\documentclass[twocolumn]{aastex701} 
\usepackage{amsmath}
\shorttitle{Infrared OH and protostellar accretion}
\shortauthors{Watson et al.}

\begin{document}

\title{IPA. Accretion rate of a low-mass Class 0 protostar,\\
measured via mid-infrared fluorescent OH emission}

\author[0000-0001-8302-0530]{Dan M. Watson}
\affiliation{Department of Physics and Astronomy, University of Rochester, Rochester, NY 14627, USA}
\email{dmw@pas.rochester.edu}

\author[0000-0002-0554-1151]{Mayank Narang}
\affiliation{Jet Propulsion Laboratory, California Institute of Technology, 4800 Oak Grove Drive, Pasadena, CA 91109, USA}
\affiliation{Academia Sinica Institute of Astronomy \& Astrophysics, Taipei 10617, TW, R.O.C.}
\email{mayank.narang@jpl.nasa.gov}

\author[0000-0001-9301-6252] {Caeley V. Pittman}
\affiliation{Department of Astronomy, Boston University, 725 Commonwealth Avenue, Boston, MA 02215, USA}
\affiliation{Institute for Astrophysical Research, Boston University, 725 Commonwealth Avenue, Boston, MA 02215, USA}
\email{cpittman@bu.edu}

\author[0000-0002-9497-8856]{Himanshu Tyagi}
\affiliation{Department of Astronomy \& Astrophysics Tata Institute of Fundamental Research, Colaba, Mumbai, Maharashtra, IN}
\email{himanshu.tyagi@tifr.res.in}

\author[0000-0002-6447-899X]{Robert Gutermuth}
\affiliation{Department of Astronomy, University of Massachusetts Amherst, Amherst, MA 01003, USA}
\email{rgutermu@astro.umass.edu}

\author[0000-0001-8790-9484]{Adam E. Rubinstein}
\affiliation{Department of Physics and Astronomy, University of Rochester, Rochester, NY 14627, USA}
\email{arubinst@ur.rochester.edu}

\author[0000-0001-5175-1777]{Neal J. Evans II}
\affiliation{Department of Astronomy, University of Texas at Austin, Austin, TX 78712, USA}
\email{nje@astro.as.utexas.edu}

\author[0000-0003-1430-8519]{Lee W. Hartmann}
\affiliation{Department of Astronomy, University of Michigan -- Ann Arbor, Ann Arbor, MI 48109, USA}
\email{lhartm@umich.edu}

\author[0000-0001-7629-3573]{S. Thomas Megeath}
\affiliation{Ritter Astrophysical Research Center, Department of Physics and Astronomy, University of Toledo, Toledo, OH 43606, USA}
\email{tommegeath@gmail.com}

\author[0000-0002-3530-304X]{P. Manoj}
\affiliation{Department of Astronomy \& Astrophysics Tata Institute of Fundamental Research, Colaba, Mumbai, Maharashtra, IN}
\email{mpuravankara@gmail.com}

\author[0000-0002-2865-7106] {Catherine C. Espaillat}
\affiliation{Department of Astronomy, Boston University, 725 Commonwealth Avenue, Boston, MA 02215, USA}
\affiliation{Institute for Astrophysical Research, Boston University, 725 Commonwealth Avenue, Boston, MA 02215, USA}
\email{cce@bu.edu}

\author[0000-0002-3950-5386]{Nuria Calvet}
\affiliation{Department of Astronomy, University of Michigan -- Ann Arbor, Ann Arbor, MI 48109, USA}
\email{ncalvet@umich.edu}

\author[0000-0001-8876-6614]{Alessio Caratti o Garatti}
\affiliation{INAF - Osservatorio Astronomico di Capodimonte, via Moiariello 16, 80131 Napoli, Italy}
\email{alessio.caratti@inaf.it}

\author[0000-0001-7591-1907]{Ewine F. van Dishoeck}
\affiliation{Leiden Observatory, Leiden University, 2300 RA Leiden, South Holland, NL}
\email{ewine@strw.leidenuniv.nl}

\author[0000-0001-7491-0048]{Tyler L. Bourke}
\affiliation{SKA Observatory, Jodrell Bank, Lower Withington, Macclesfield SK11 9FT, UK}
\email{tyler.bourke@skao.int}

\author[0000-0003-1665-5709]{Joel D. Green}
\affiliation{Space Telescope Science Institute, Baltimore, MD 21218, USA}
\email{jgreen@stsci.edu}

\author[0000-0002-9548-1526] {Carey M. Lisse}
\affiliation{Planetary Exploration Group, Space Department, Johns Hopkins University Applied Physics Laboratory, Laurel, MD 2072}
\email{carey.lisse@jhuapl.edu}

\author[0000-0001-9443-0463]{Pamela Klaassen}
\affiliation{United Kingdom Astronomy Technology Centre, Royal Observatory Edinburgh, Blackford Hill, Edinburgh EH9 3HJ, UK}
\email{pamela.klaassen@stfc.ac.uk}

\author[0000-0002-4540-6587]{Leslie W. Looney}
\affiliation{Department of Astronomy, University of Illinois, Urbana, IL 61801, USA}
\email{lwl@illinois.edu}

\author[0000-0002-4448-3871]{Pooneh Nazari}
\affiliation{Leiden Observatory, Leiden University, 2300 RA Leiden, South Holland, NL}
\affiliation{European Southern Observatory, 85748 Garching, DE}
\email{pooneh.nazari@eso.org}

\author[0000-0001-8341-1646]{David A. Neufeld}
\affiliation{William H. Miller III Department of Physics and Astronomy, The Johns Hopkins University, Baltimore, MD, USA}
\email{neufeld@jhu.edu}

\author[0000-0002-6195-0152]{John J. Tobin}
\affiliation{National Radio Astronomy Observatory, Charlottesville, VA 22903, USA}
\email{jtobin@nrao.edu}

\author[0000-0002-0826-9261]{Scott J. Wolk}
\affiliation{Harvard-Smithsonian Center for Astrophysics, Cambridge, MA 02138, USA}
\email{swolk@cfa.harvard.edu}

\author[0000-0002-7506-5429]{Guillem Anglada}
\affiliation{Instituto de Astrof{\'i}sica de Andaluc{\'i}a, CSIC, Glorieta de la
Astronom{\'i}a s/n, E-18008 Granada, ES}
\email{guillem@iaa.es}

\author[0000-0002-4026-126X]{Prabhani Atnagulov}
\affiliation{Ritter Astrophysical Research Center, Dept. of Physics and Astronomy, University of Toledo, Toledo, OH, USA}
\email{prabhani.rajakaruna@rockets.utoledo.edu}

\author[0000-0002-1700-090X]{Henrik Beuther}
\affiliation{Max Planck Institute for Astronomy, Heidelberg, Baden-Württemberg, DE}
\email{beuther@mpia-hd.mpg.de}

\author[0000-0001-7826-7934]{Nashanty G. C. Brunken}
\affiliation{Leiden Observatory, Leiden University, 2300 RA Leiden, South Holland, NL}
\email{brunken@strw.leidenuniv.nl}

\author[0000-0002-6136-5578]{Samuel Federman}
\affiliation{Ritter Astrophysical Research Center, Department of Physics and Astronomy, University of Toledo, Toledo, OH 43606, USA}
\affiliation{INAF-Osservatorio Astronomico di Capodimonte, IT}
\email{sam.federman@gmail.com}

\author[0000-0001-9800-6248]{Elise Furlan}
\affiliation{Caltech/IPAC-NASA Exoplanet Science Institute, Pasadena, CA 91106, USA}
\email{furlan@ipac.caltech.edu}

\author[0000-0002-2667-1676]{Nolan Habel}
\affiliation{Jet Propulsion Laboratory, California Institute of Technology, 4800 Oak Grove Drive, Pasadena, CA 91109, USA}
\email{nmhabel@gmail.com}

\author[0000-0003-3682-854X]{Nicole Karnath}
\affiliation{Space Science Institute, Boulder, CO 80301, USA}
\email{nkarnath@gmail.com}

\author[0000-0002-5943-1222]{Hendrik Linz}
\affiliation{Max Planck Institute for Astronomy, Heidelberg, Baden-Württemberg, DE}
\email{linz@mpia.de}

\author[0000-0002-5943-1222]{James Muzerolle Page}
\affiliation{Space Telescope Science Institute, Baltimore, MD 21218, USA}
\email{muzerol@stsci.edu}

\author[0000-0002-6737-5267]{Mayra Osorio}
\affiliation{Instituto de Astrof{\'i}sica de Andaluc{\'i}a, CSIC, Glorieta de la Astronom{\'i}a s/n, E-18008 Granada, ES}
\email{osorio@iaa.es}

\author[0000-0002-0557-7349]{Riwaj Pokhrel}
\affiliation{Ritter Astrophysical Research Center, Dept. of Physics and Astronomy, University of Toledo, Toledo, OH 43606, USA }
\email{riwajpokhrel@gmail.com}

\author[0000-0002-5350-0282]{Rohan Rahatgaonkar}
\affiliation{Gemini South Observatory, La Serena, CL}
\email{rohan.rahatgaonkar@rockets.utoledo.edu}

\author[0000-0001-6144-4113]{Will R. M. Rocha}
\affiliation{Leiden Observatory, Leiden University, 2300 RA Leiden, South Holland, NL}
\email{rocha@strw.leidenuniv.nl}

\author[0000-0002-9209-8708]{Patrick D. Sheehan}
\affil{National Radio Astronomy Observatory, Charlottesville, VA 22903 USA} 
\email{psheehan@nrao.edu}

\author[0000-0002-7433-1035]{Katerina Slavicinska}
\affiliation{Leiden Observatory, Leiden University, 2300 RA Leiden, South Holland, NL}
\email{slavicinska@strw.leidenuniv.nl}

\author[0000-0002-5812-9232]{Thomas Stanke}
\affiliation{Max Planck Institute for Extraterrestrial Physics, Gie\ss{}enbachstraße 185748 Garching, DE}
\email{tstanke049@gmail.com}

\author[0000-0003-2300-8200]{Amelia M.\ Stutz}
\affiliation{Departamento de Astronom\'{i}a, Universidad de Concepci\'{o}n, Concepci\'{o}n, CL}
\email{astutz@astro-udec.cl}

\author[0000-0002-9470-2358]{Lukasz Tychoniec}
\affiliation{European Southern Observatory, Garching bei M\"unchen, DE}
\email{lukasz.tychoniec@gmail.com}

\author[0000-0001-8227-2816]{Yao-Lun Yang}
\affiliation{RIKEN Cluster for Pioneering Research, Wako-shi, Saitama, 351-0106, JP}
\email{yaolunyang.astro@gmail.com}

\author[0000-0002-3747-2496]{William J. Fischer}
\affiliation{Space Telescope Science Institute, Baltimore, MD 21218, USA}
\email{wfischer@stsci.edu}

%\correspondingauthor{Dan Watson}

\begin{abstract}
The earliest stages of star formation are highlighted by complex interactions between accretion, outflow, and radiative processes, which shape the chemical and physical environment of the emerging protostar. James Webb Space Telescope observations of the low-mass, low-luminosity Class 0 protostar IRAS 16253-2429 reveal a central compact source.  This object exhibits a rich mid-infrared emission spectrum of OH pure rotational lines and $\rm CO_2$ ro-vibrational lines.  Unusually for a young stellar object, it has no mid-infrared line emission from $\rm H_2O$ to match the other molecules. We demonstrate that the emitting OH molecules arise from UV photodissociation of $\rm H_2O$ in its second absorption band at $\lambda = 114-145$ nm, and that the OH emission is a fluorescent cascade starting with highest-excitation rotational states.  This situation offers the opportunity of using the infrared OH spectrum to measure the UV flux from the central protostar.  Thereby we determine the disk-star accretion rate to be $3
\times 10^{-10} \ M_\sun \ {\rm year^{-1}}$,  and demonstrate that the system luminosity arises mostly from the protostar's photosphere rather than from accretion luminosity. The result is in accord with the measured outflow rate of IRAS 16253-2429 and lies within the  outflow/accretion-flow rate trend often inferred for protostars; and with episodic accretion as the dominant mechanism by which this protostar has grown.
\end{abstract}


\section{Dedication} 
    \label{will}
    We dedicate this article to the memory of our colleague William J. Fischer, who passed away tragically on 16 April 2024, in the prime of his life and career.  Will was well known for his work on star formation, the evolution of young stars, and especially for infrared and UV studies of accretion processes.  Most of us have worked closely with Will for more than a decade; he was a great teammate and a great friend, and his death is a profound loss.

\section{Introduction} 
    \label{intro}
    Mass assembly in low-mass protostars is thought to occur via the infall of gas onto a disk and the subsequent accretion of that gas from the disk to star (e.g. \citealt{Hartmann+2016}). Measurements of the disk to star accretion rate, $\dot{M}_a$, are therefore fundamental to studies of protostellar evolution. Among several ways to measure $\dot{M}_a$, the best is {\it via} ultraviolet excess: namely, continuum and line emission in excess of the protostar's photosphere (e.g., \citealt{Hartmann_2009}). The excess arises mostly in shocked gas, energized by impact of disk material as it falls from several stellar radii away (\citealt{Calvet+1998}, \citealt{pittman+2025}). The main difficulty in direct measurement of UV excess is extinction in the foreground of the star. This hindrance is merely significant for Class II young stellar objects (YSOs), but becomes insuperable for the spectral-energy-distribution classes (0, I, flat-spectrum) characteristic of early evolutionary stages. Supplementary methods, based on empirical relationships between near-infrared H I recombination line fluxes and accretion luminosity (e.g. \citealt{Muzerolle+1998}, \citealt{alcala+2014}, \citealt{alacala+2017}, \citealt{Komarova+2020}, \citealt{Fiorellino+2023}, \citealt{Tofflemire+2025}, \citealt{testi+2025}), extend our grasp into this envelope-dominated range, though they are difficult to use if envelope extinction is very large \citep{Gouellec+2024}, if accretion occurs through a disk boundary layer (\citealt{Hartmann+2016}, \citealt{Fischer+2012}), or if substantial H I emission arises from processes related to outflows (e.g. \citealt{Narang+2024}, \citealt{Federman+2024}). 

Here we present a different method, of promise for low-mass YSOs in these early stages: photodissociation of water in the inner regions of protostars, within range of the star's accretion-generated ultraviolet.  Hydroxyl (OH) radicals produced thereby emit a fluorescent cascade in the mid infrared ($\lambda > 5~\micron$). The OH emission is directly dependent upon the UV luminosity and the rate of mass accretion that produced it. 

We apply this novel method to IRAS 16253-2429 (henceforth I16253), a Class 0 protostar in the main Ophiuchus/L1688 molecular cloud.  I16253's bolometric luminosity is $L_{bol} = 0.16 \ L_\sun$ at distance $d_0 = 140 \ \rm{pc}$ (\citealt{Aso+2023}, \citealt{Pokhrel+2023}).  The central star's mass is determined kinematically to be $M = 0.15\pm 0.03 \ M_\sun$ (\citealt{Aso+2023}, \citealt{Hsieh+2019}). Its protostellar envelope is large by comparison, at $M_{env} = 1.0 \pm 0.5 M_\sun$ (\citealt{Young+2006}, \citealt{Tobin+2011}), and infall motions are detected in the envelope's inner parts \citep{Aso+2023}. It is the source of a bipolar molecular outflow with length $\ell = 0.11~{\rm pc}$ and dynamical age $\Delta t \sim 10^4 ~{\rm yr}$ (e.g. \citealt{Hsieh+2017}, \citealt{Barsony+2010}). I16253's mid-infrared spectrum contains many spectral lines of OH, which are bright compared to those of the species which usually dominate the spectra of YSOs.  In the following we show that water surrounding I16253 is photodissociated by protostellar UV into OH and atomic hydrogen. At I16253's mass and luminosity, the necessary UV emission must arise in disk-star accretion flows. 

\section{Observations and data reduction}   
    \label{Observations}
    \begin{figure*}
    \includegraphics[width=1.0\textwidth]{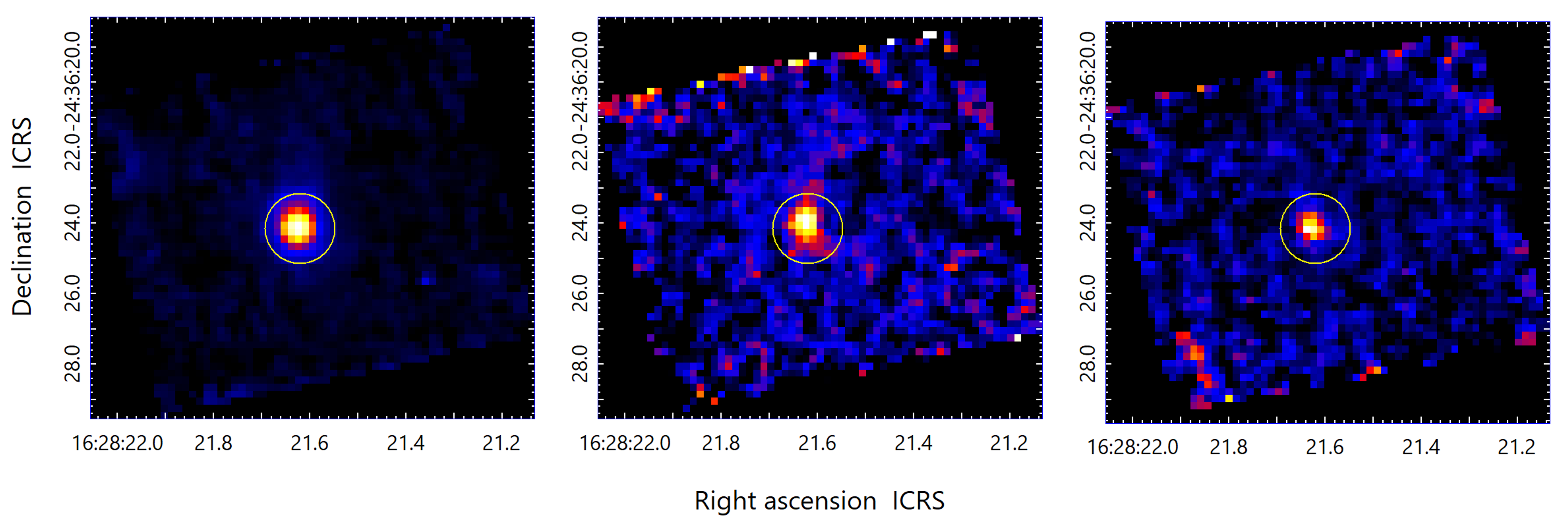}
    \caption{
    Images of the central object in IRAS 16253-2429.  From left: $\lambda~=~14~\micron$ continuum, arcsinh stretch; $\rm{CO_2}\ 0 1^1 0 - 0 0^0 0$ Q branch, $\lambda = 14.98~\micron$, linear stretch; and OH X1.5 $N = 16~A'$, $\lambda = 16.84~\micron$, linear stretch.  Brightness limits are $-1\sigma$ (black) to peak intensity (white) in each case. The yellow circle is the 1 arcsec radius extraction aperture for the measurements listed in Table \ref{list_of_fluxes}, and is centered on ICRS coordinates $\alpha = 16:28:21.60$, $\delta = -24:36:23.4$ \citep{Hsieh+2017}.    
    \label{Images}}
\end{figure*}
\begin{figure*}
    \includegraphics[width=1.0\textwidth]{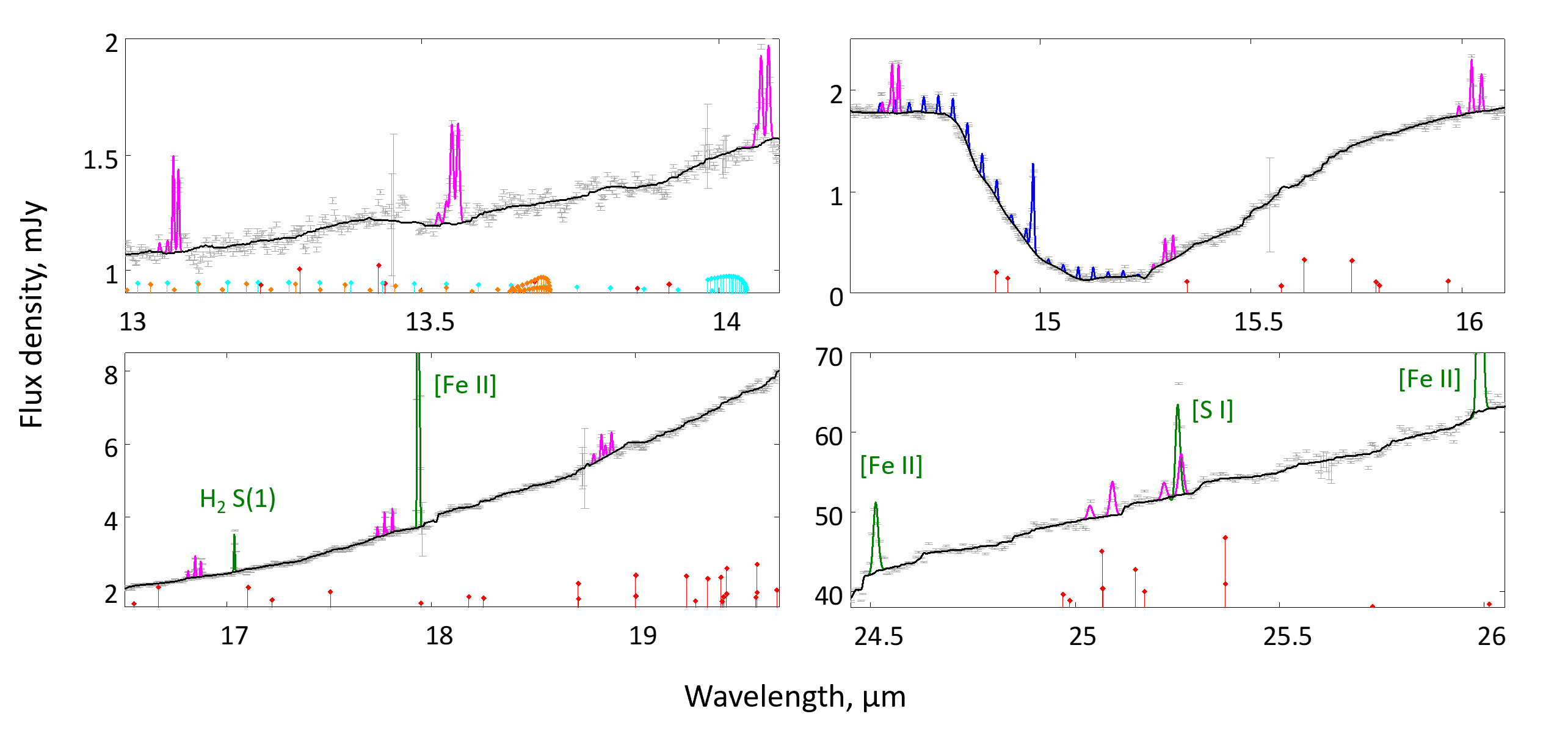}
    \caption{
    Representative sections of the JWST-MIRI spectrum of I16253, shown as gray points and errorbars. The latter include propagated noise and flux-calibration uncertainty. The continuum fit is overlaid in black, along with fits of the MIRI spectral point-spread function to emission lines of OH (pink), CO$_2$ (blue), and assorted H$_2$ and atomic or ionic lines associated with the protostar’s outflow: collimated jet or wider-angle wind, or cavity wall (green). Marked along the bottom of each frame are lines not detected: H$_2$O in red, HCN in cyan, and C$_2$H$_2$ in orange. They are plotted with notional flux densities, in the vertical scale of their frame and measured from the wavelength axis. These flux densities are calculated from the relative abundances and excitation temperatures found for these molecules in the Class II YSO GW Lup by \cite{Grant+2023}, which we use only as an example of a spectrum containing the complete cast of strong mid-infrared lines from YSOs. The number of $\rm CO_2$ molecules we detect is an order of magnitude larger than detected in GW Lup, but the temperature we find for them $(T = 110~\rm K)$ is substantially smaller than GW Lup's $T$ = 400 K (Section \ref{extinction}).    
    \label{Spectra}}
\end{figure*}

We observed I16253 on 2022 July 22-23, as part of the James Webb Space Telescope (JWST) General Observing Program 1802 ({\it Investigating Protostellar Accretion} [IPA]; S.T. Megeath, PI).  We used the JWST Mid-Infrared Instrument (MIRI) Medium-Resolution Spectrograph (MRS) integral-field-unit (IFU), to obtain spectral images at wavelengths $\lambda~=~4.9-27~\micron$ of 6 arcsec $\times$ 6 arcsec to 9 arcsec $\times$ 9 arcsec fields centered on the protostar.  The spectral images were comprised of $2\times 2$ mosaics of individual IFU frames with at least 10\% overlap, each observed with four-point dithers. The MIRI MRS IFU pixel scale is $0.27\times 0.27~\rm{arcsec}^2$ at its longest wavelengths and $0.2\times 0.2~\rm{arcsec}^2$ at its shortest.  Thus the diffraction-limited point spread function, of diameter 1 arcsec$\times\lambda /27~\micron$, is well sampled by MIRI's IFU for $\lambda\gtrsim 10~\micron$. We used all twelve MIRI MRS spectral sub-bands, obtaining resolving power ranging over $\lambda/\Delta\lambda~=~1350-3700$ as wavelength ranges over $\lambda~=~27-4.9~\micron$.  More details of our observations are given by \cite{Narang+2024}, \cite{Rubinstein+2024}, and \cite{Federman+2024}.

\section{Data analysis}
    \label{analysis}
    \begin{figure*}
    \includegraphics[width=0.9\textwidth]{energy_levels_2.png}
    \caption{
   Energy level diagram of the lowest-energy v = 0 states of OH. Left: rotational states of the ground electronic configuration, with fine structure and $\Lambda$-doubling shown, and detected transitions indicated with pink lines. Inset: example MIRI spectrum of the $N$ = 16-15 transitions. Right: zoom in to the states involved in the $N$ = 16-15 transitions. For clarity, the $\Lambda$-doublet $A''$ and $A'$ states are each plotted with symmetrical displacement about their total angular momentum quantum number $J$. Comparison of the inset spectrum and the zoomed view illustrates the faintness of the $A''$ lines relative to the $A'$ transitions, a common feature of the OH lines in Figure \ref{Spectra}.
    \label{energy_levels}}
\end{figure*}

Coincident with the protostellar position in our data cubes is an unresolved (${\rm FWHM} < 80~\rm{AU}$ diameter) source of spectral lines and continuum at $\lambda\geq 10~\micron$, shown in Figure \ref{Images}. We chose an aperture with radius 1 arcsecond, large enough to encompass this source at the longest MIRI wavelengths. Within this aperture we extracted the flux density $F_\nu$ at $\lambda = 4.9-27~\micron$ from the data cubes. We masked wavelengths at and around those of spectral lines we detected; fit the remaining baseline, smoothed over 51 channels to a cubic spline; subtracted this baseline from the data; and simultaneously fit Gaussian profiles with the instrumental-resolution width to the baseline-subtracted data at the wavelengths of each spectral line.  For ionic, atomic, and $\rm H_2$ lines, which arise in I16253's high-speed jet or wider-angle wind, \citep{Narang+2024}, we included several velocity components to ensure capture of all the flux.  All other lines we detect are spectrally unresolved and consistent with the protostar's systemic radial velocity. Table \ref{list_of_fluxes} is a list of the emission lines we detect, and upper limits for the OH pure rotational lines we do not.  In Figure \ref{Spectra} we show example segments of the extracted spectrum. 

I16253's mid-infrared emission-line spectrum is dominated by lines of OH and gas-phase $\rm CO_2$. We detect every pure rotational transition of OH in the $\lambda~=~10-27~\micron$ range: both of the $X^2\Pi_{3/2}$ and $X^2\Pi_{1/2}$ electronic configurations (henceforth X1.5 and X0.5); lower-state rotational quantum numbers $N = 10-33$; and generally both $\Lambda$-doublet components $A''$ and $A'$ resolved from one another, as in Figures \ref{Spectra} and \ref{energy_levels}.  

We also detect the P, Q, and R branches (respectively $\Delta J = +1, 0, -1$) of the fundamental, $\nu_1 \nu_2^\ell \nu_3 = 0 1^1 0 - 0 0^0 0$ rotation-vibration band of $\rm CO_2$.  The rotational structure of the Q branch is spectrally unresolved, but the P and R branches are distinct, and the latter is detected through R(16) (i.e. $J = 17-16$).  In Appendix \ref{Appendix} we briefly summarize the mid-infrared spectroscopy of OH and $\rm CO_2$, and define the terms we use here.  The unresolved nature of I16253's OH and $\rm CO_2$ emission (Figure \ref{Images}) suggests that it arises near $(< 40~\rm AU)$ the central protostar.  

This collection of lines is unlike those of most low-mass YSOs, Class 0/I \citep{vanGelder+2024a} or Class II (e.g. \citealt{Pontoppidan+2010}, \citealt{Carr+2011}).  We do not detect several molecular species whose lines usually dominate the mid-infrared spectra of protoplanetary disks, notably $\rm{H_2O}$, HCN, and $\rm{C_2H_2}$. Upper limits on $\lambda~=~10-27~\micron$ lines of these species are much smaller than those of the detected OH and $\rm CO_2$ lines. Most Class 0/I objects observed so far display lines of both $\rm H_2O$ and $\rm CO_2$, with the former brighter than the latter \citep{vanGelder+2024a}. In Figure \ref{Spectra} we mark the lines of these undetected molecules and indicate their relative strength in a Class II YSO spectrum which contains all these molecular spectra.  In I16253 we also do not detect vibrationally-excited OH, higher-energy ro-vibrational modes of $\rm{CO_2}$, or rotational transitions between the OH X1.5 and X0.5 ladders.  

I16253's $\lambda = 15~\micron$ $\rm CO_2$ ice absorption feature turns out to yield useful constraints on the nature of the OH and $\rm CO_2$ line emission.  We show a line-subtracted spectrum of the ice feature's optical depth $\tau$ in Figure \ref{ice_spectrum}.  Its shape is similar to that toward many other Class 0/I objects which have been modeled as cold $(T \lesssim 35~\rm K)$ icy grains along the line of sight through the protostar's envelope (e.g. \cite{Pontoppidan+2008}, \cite{Zasowski+2009}); it is similar to the ice feature recently observed toward the edge-on Class 0 object Ced 110 IRS4A \citep{Rocha+2025}, to which the embedded disk probably makes significant contributions. It lacks very prominent segregated-$\rm CO_2$ features at $\lambda = 15.1$ and $15.25~\micron$, as in objects which have warmer $(> 80\rm K)$ ices \citep{Brunken+2025}.  We modeled the absorption with cold $(T = 10~\rm K)$, small $(a \ll \lambda)$ grains of a continuous distribution of ellipsoidal shapes (CDE; \cite{Bohren+1998}, equation 12.36), using optical constants in the Leiden Ice Database for Astrophysics\footnote{\label{leiden_ice} See https://icedb.strw.leidenuniv.nl/.}, 
and CDE grain absorbance from \cite{Rocha+2025} and \cite{Pontoppidan+2008}.  From the fitted components we computed molecular column densities $\mathcal{N}$,

\begin{equation}
   \mathcal{N} = \frac{1}{A_b}\int \tau d\nu~~~,
\label{ice}
\end{equation}

\noindent and took the band strength $A_b$ to be $1.1\times 10^{-17}~\rm cm~molecule^{-1}$ \citep{Pontoppidan+2008}.  The resulting minimum-$\chi^2$ fit, shown in Figure \ref{ice_spectrum}, is similar to the spectral decomposition of Ced 110 IRS4A by \cite{Rocha+2025}.  We find a column density ratio $\mathcal{N}(\rm{H_2O})/\mathcal{N}(\rm{CO_2}) = 1.9\pm 0.7$, placing the object along one edge of the distribution of this ratio in the survey by \cite{Pontoppidan+2008}, but not discordant with the population.  We will assume, below (section \ref{accretion}), that all significant components of ice in this young protostar have this same value of $\mathcal{N}(\rm{H_2O})/\mathcal{N}(\rm{CO_2})$, though because a relatively narrow selection of optical constants was used, and the ratio is expected to vary from place to place within I16253, we will not propagate the fitting uncertainties in the calculations below.

\section{Results} 

\subsection{I16253's OH emission is fluorescence driven by accretion-generated UV}
    \label{fluorescence}
    I16253's OH emission is predominantly from lines between the lower-energy, $A'$ members of $\Lambda$-doublets, equally bright in the X1.5 and X0.5 configurations. The $A''$ lines are much fainter, especially for larger values of $N$. This is apparent in Figures \ref{Spectra} and \ref{energy_levels}, and in the flux ratios for the full range of $N$ as plotted in Figure \ref{pop_ratios}.  However, the radiative decay rates (Einstein A coefficients) of $A''$ and $A'$ states are practically the same, for a given $N$ (\citealt{Brooke+2016}, \citealt{Noll+2020})\footnote{See https://hitran.org/.}.  Taken together with the small $A''/A'$ flux ratios, this requires all the OH lines to be optically thin, with flux directly proportional to the number of upper-state molecules. Thus the populations of the $\Lambda$-doublet states are strongly anti-inverted.  This condition cannot be produced or significantly modified by collisional processes.  Collisional de-excitation rates have not been calculated for OH rotational states in the present range, but those at smaller $N$ (\citealt{Offer+1994}, \citealt{Kalugina+2014}) suggest that collisions would lead to $A''$ and $A'$ state populations closely similar to one another. 

Photodissociation of $\rm H_2O$ by photons in water's second absorption band (SAB) at $\lambda = 114-145~\rm nm$ efficiently produces OH with properties precisely replicating our observations.  In laboratory measurements and simulations alike (\citealt{Mordaunt+1994}, \citealt{Dixon_1995}, \citealt{Harich+2000}, \citealt{Zhou+2015}, \citealt{Young+2018}, \citealt{Chang+2019}), $\lambda = 121.6~\rm{nm}$ photodissociation of $\rm{H_2O}$ produces OH with low electronic and vibrational excitation: about 82\% in X1.5 and X0.5, in equal numbers, and 75\% in $v = 0$.  The products have high rotational excitation, $N = 30-45$. If the target $\rm{H_2O}$ is rotationally cold ($T \lesssim 200~K$), a large fraction of the OH is produced with $A''/A' \ll 1$, and with $A''/A' \rightarrow 0$ as $T \rightarrow 0$ \citep{Dixon_1995}. 

As these OH molecules are produced with large values of $N$, whether in the ground electronic and vibrational manifolds or not, they decay rapidly to other large-$N$ states of $X {}^2\Pi$ $v = 0$.  From there, they decay down the rotational ladders of X1.5 and X0.5. All the states have small lifetimes, 0.001-0.01 sec for $N = 10-33$, and
the molecules reach the lower ($N < 10$) rotational states before they are themselves dissociated by protostellar UV ($> 10^4~\rm{sec}$; Section \ref{accretion}).  Thus the steady-state populations of the OH rotational states are given in terms of the state populations and their A coefficients as 

\begin{equation}
    n_{J+1}A_{J+1,J} = n_{J}A_{J,J-1}~~~,
\label{cascade}
\end{equation}

\noindent
each term obtained from an OH rotational line flux.  The sum of the four values of $n_{J}A_{J,J-1}$ which belong to orbital angular momentum state $N$ is $\dot{n}_{\rm f}(N)$, the rate in photons $\rm sec^{-1}$ at which OH molecules pass through the fluorescent cascade.  According to Equation \ref{cascade}, $\dot{n}_{\rm f}(N)$ is independent of $N$ in a steady state.

\begin{figure}
    \includegraphics[width=0.47\textwidth]{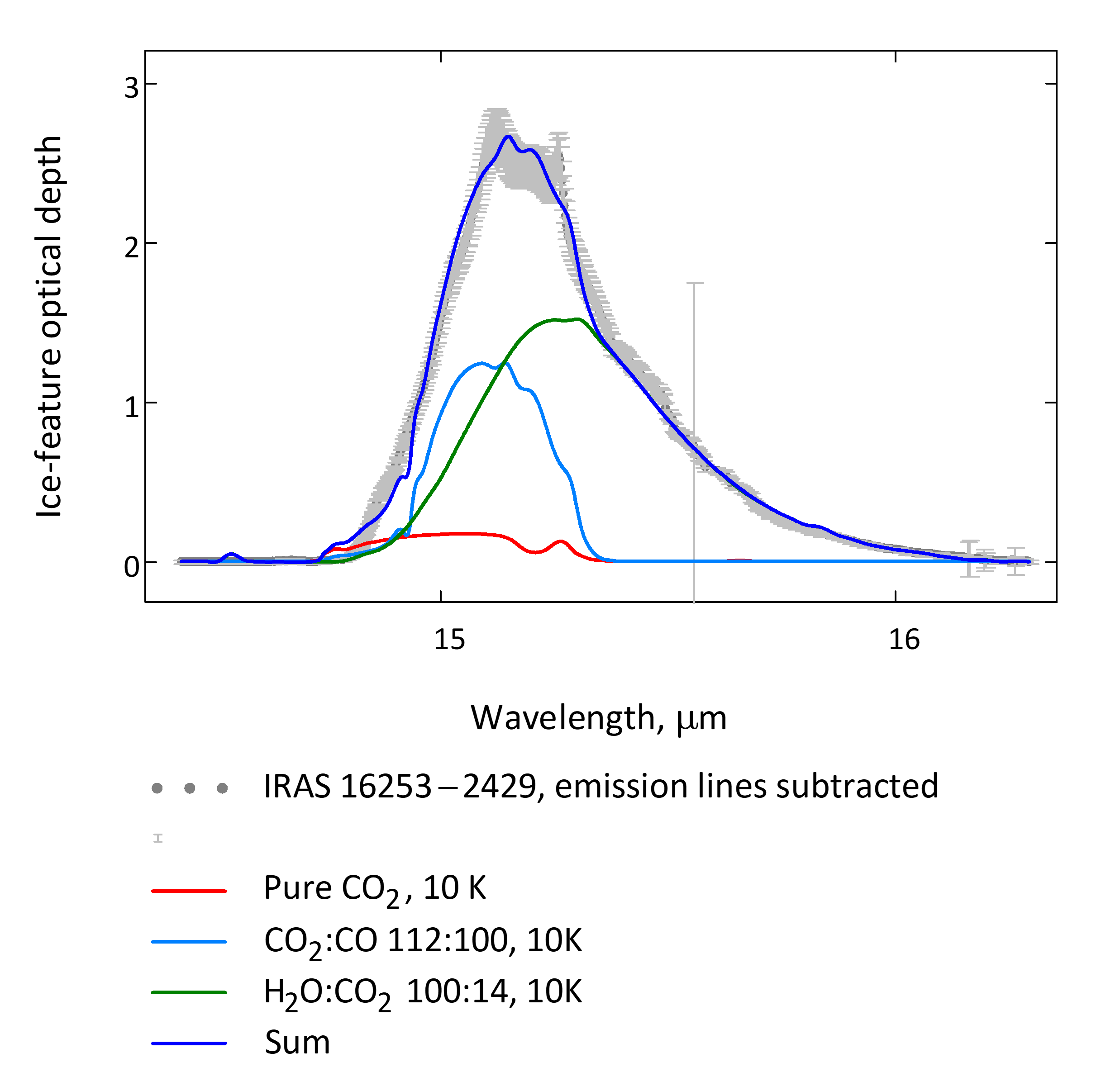}
    \caption{
    Optical depth spectrum of the $\rm{CO_2}$ ice absorption feature.  Overlaid on the JWST data (gray points and errorbars) is the best fit found for small ($\ll 15~\micron$) grains of CDE shape distribution, coated with cold $(T = 10~\rm K)$ ice analogues: $\rm{CO_2:CO}$ from \cite{Pontoppidan+2008} and \cite{Rocha+2025},  $\rm{H_2O:CO_2}$ and pure $\rm{CO_2}$ from \cite{Ehrenfreund+1997}. This ice mixture has column density ratio $\mathcal{N}(\rm{H_2O})/\mathcal{N}(\rm{CO_2}) = 1.9$.
    \label{ice_spectrum}}
\end{figure}
\begin{figure}
    \includegraphics[width=0.47\textwidth]{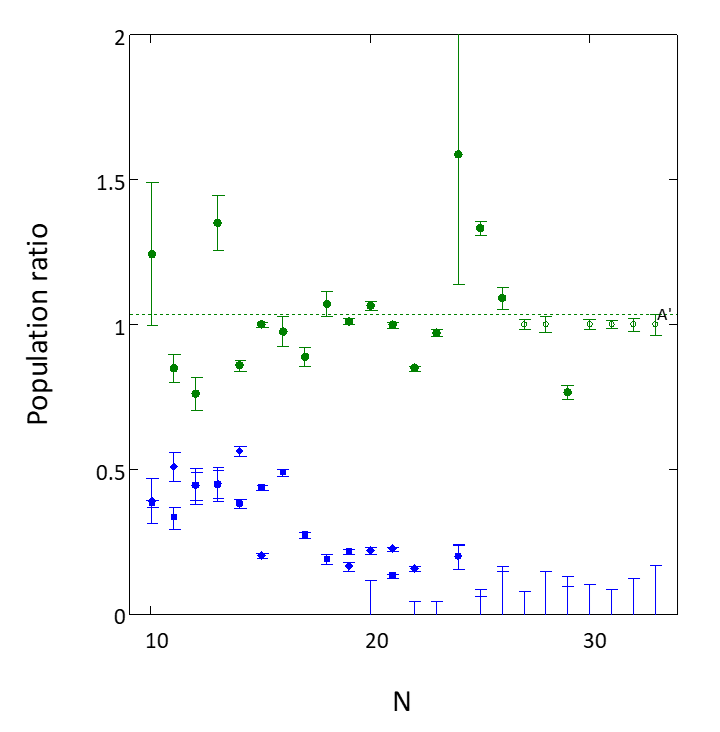}
    \caption{
    OH rotational-state population ratios: X1.5 $A^{\prime\prime}/A^{\prime}$ (blue squares), X0.5 $A^{\prime\prime}/A^{\prime}$ (blue diamonds), and $X0.5~A^\prime /X1.5~A^\prime$ (green circles), all as functions of the lower-state orbital angular momentum quantum number $N$.  As these ratios are extremely insensitive to extinction, they are plotted without extinction correction. The open green circles at $N$ = 27-28 and 30-33 represent spectrally-unresolved combinations of $X0.5~A^\prime /X1.5~A^\prime$ lines, plotted with notional ratios of unity merely to show their propagated uncertainties. The upper limits at $N = 27-28$ and $30-33$ account for both $A^{\prime\prime}/A^{\prime}$ ratios and the unresolved $A^\prime -A^\prime$ pair.  The dashed green line marks the mean value of the solid green points, $1.04\pm 0.21$. 
    \label{pop_ratios}}
\end{figure}

The OH $\Lambda$-doublets have anti-inverted populations in all the rotational states we observe. However, $A''/A'$ increases as $N$ decreases, reaching $A''/A' = 0.4-0.5$ at $N \lesssim 15$ (Figure \ref{pop_ratios}).  This may indicate contribution to the OH spectrum in this range by processes besides fluorescence, as we address below. 

A small number of previous YSO-related observations reveal prominence of OH emission over $\rm{H_2O}$ emission similar to our spectra.  \cite{Tappe+2008}, \cite{Tappe+2012}, and \cite{Tabone+2021} invoked SAB photodissociation of $\rm H_2O$ in their account of OH spectra of terminal shocks in YSO outflows. So did \cite{Neufeld+2024}, with JWST-MIRI spectra of HOPS 370's jet.  The mid-infrared line spectrum of the Class II YSO TW Hya is similar to that of I16253; \cite{Najita+2010} invoked SAB photodissociation of $\rm H_2O$ and prompt emission by OH to explain the OH emission.  \cite{Carr+2011} and \cite{Carr+2014} ascribed the OH emission from the Class I/II YSO DG Tau in part to SAB photodissociation of $\rm H_2O$ followed by OH rotational-line fluorescence.  Recently the Class II YSO CX Tau was also suggested to have UV-excited fluorescent OH rotational emission \citep{Vlasblom+2025}.  \cite{Zannese+2024} invoked photodissociation to explain the OH emission from externally-illuminated protoplanetary disks adjacent to the Orion Nebula. In solar-system comets, \cite{Bonev+2006} observed OH population ratios in $v$ = 1, $N = 16$ similar to those we see in the same rotational states of $v$ = 0 (Table \ref{list_of_fluxes}).  They suggested SAB photodissociation products as the origin, among other possibilities.   

Three alternatives to OH production by SAB dissociation of $\rm H_2O$ are not supported by our observations: $\rm H_2O$ photodissociation in $\rm H_2O$'s first absorption band (FAB; $\lambda = 145-186$ nm); OH production by $\rm O - H_2$ reaction; and collisional and radiative excitation by warm gas and/or the protostar's infrared continuum emission:

\begin{itemize}
    \item $\rm{H_2O}$ photodissociation in the FAB leads to OH predominantly in the X1.5 and X0.5 configurations, but in $v\geq 1$ vibrational states and $N\lesssim 5$ rotational states, and with $\Lambda$-doublet populations inverted: $A''/A' > 1$ (e.g. \citealt{Andresen+1984}): the diametric opposite of our observations. In obedience to the selection rules, OH produced in this manner decays rapidly to small-$N$ rotational states within the $v = 0$ vibrational manifold, skipping over the states probed by our measurements.
    \item ${\rm O}(^1D)\rm + H_2 \rightarrow H + OH$ also produces OH in the $X ^2\Pi$ configurations, and with a wide range of rotational states populated, but also with $v \geq 1$ vibrational manifolds populated similarly to $v = 0$ (\citealt{Liu+2000}, \citealt{Alexander+2004}). This would produce many vibrationally-excited OH lines in our spectrum similar in flux to the pure rotational lines \citep{Tabone+2021}, which we do not detect.  The reaction does, however, produce $A''/A' < 1$ in the range we see, at $N \leq 15$ (\citealt{Alexander+2004}, \citealt{Tabone+2024}). Judging from the excitation and output spectrum in the Tabone et al. results, and provided that sufficient ${\rm O}(^1D)$ is present, this mechanism may contribute to the rise in $A''/A'$ and OH rotational line flux we see at the smaller values of $N$.
    \item As mentioned above, collisional excitation cannot produce small $A''/A'$. Nor can infrared radiative pumping: because the Einstein A coefficients are practically the same for a given $N$, so are the Einstein B coefficients, and the $\Lambda$ doublets are too closely spaced for a broadband source of radiation to select one $\Lambda$-doublet state over the other. 
\end{itemize}

\subsection{Accretion-generated UV from low-mass protostars} 
    \label{accretion}  
    \begin{figure}
    \includegraphics[width=0.47\textwidth]{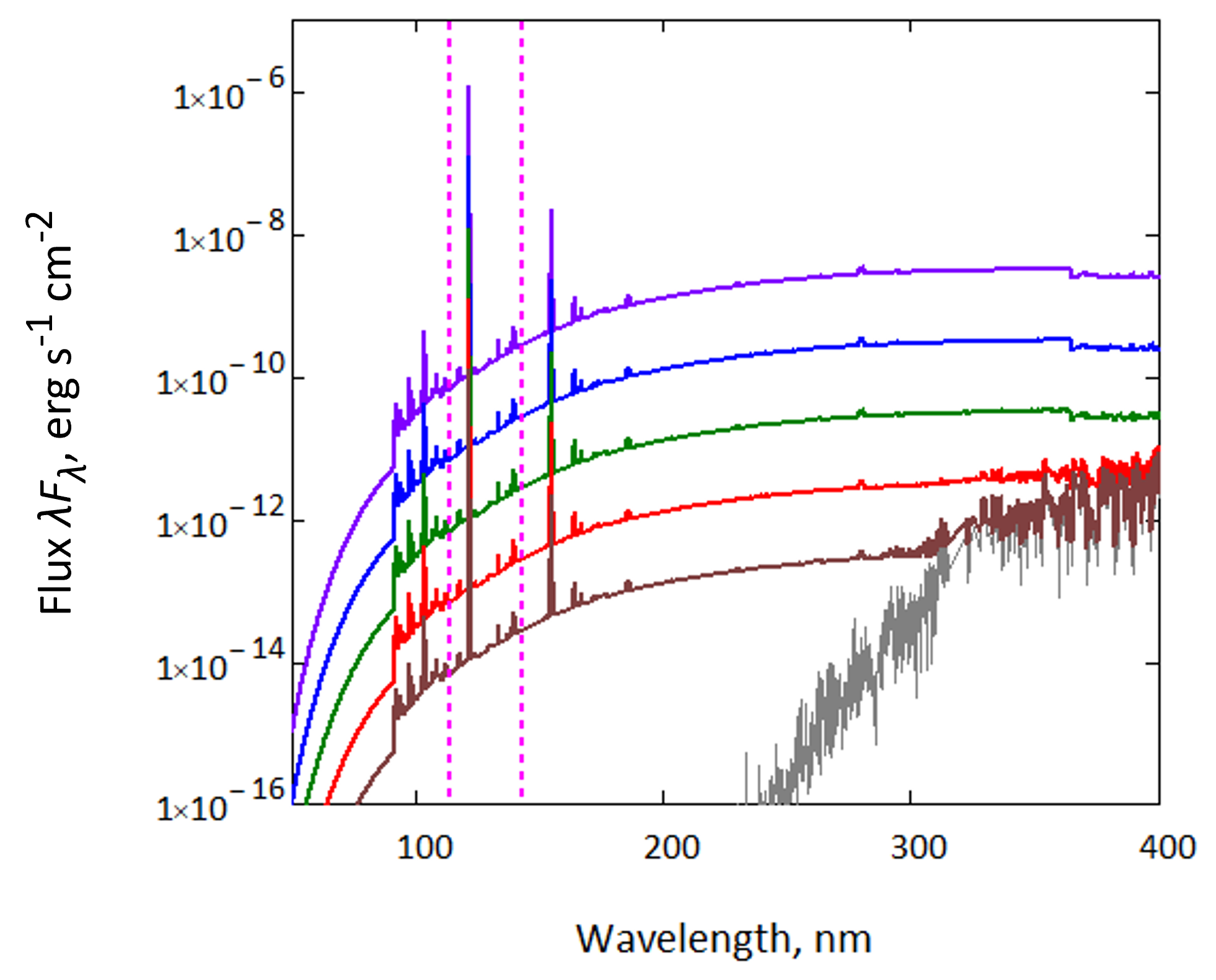}
    \caption{
    Flux $\lambda F_\lambda$ at distance $d = 140~\rm pc$, as a function of wavelength, for the small grid of accretion models described in Section \ref{accretion}. The underlying stellar spectrum is plotted in gray. Stellar plus pre- and post-shock accretion-powered flux is plotted for accretion rates $\dot{M} = 10^{-6}$ (uppermost, purple), $10^{-7}$ (blue), $10^{-8}$ (green),$10^{-9}$ (red), and $10^{-10}~M_\sun~{\rm year}^{-1}$ (brown, lowest).  Dashed pink lines mark the edges of water's second absorption band (SAB).
    \label{accretion_models}}
\end{figure}

Low-mass protostars like I16253 only produce significant UV radiation {\it via} accretion. The main alternative would be strong shocks within the protostar's high-speed jet.  However, the shocks in I16253's jet are relatively slow ($v_S \lesssim 55~\rm km~sec^{-1}$; \citealt{Narang+2024}), incapable of generating substantial UV (compare \cite{Neufeld+2024}).  Shocks within the jet also occupy a small solid angle in the view of the circum-protostellar regions.  We conclude that the UV responsible for OH production is generated by accretion processes around the protostar. 

To relate the extinction-corrected OH fluorescence rate $\langle{\dot{n}_{\rm f}}\rangle$ to the disk-star accretion rate $\dot{M}_a$, we have created a small grid of model of far-UV, $\lambda = 50~{\rm nm} - 400~{\rm nm}$, spectra (Figure \ref{accretion_models}).  The underlying stellar spectrum is that of a Phoenix model photosphere \citep{Husser+2013} for a star with mass $M = 0.15~M_\sun$, radius $R = 1.3~R_\sun$, and effective temperature $T_e = 3200 ~{\rm K}$ at distance $d_0 = 140~{\rm pc}$. The model grid covers accretion rates $\dot{M}_a = 10^{-10} - 10^{-6} ~M_\sun ~ {\rm year^{-1}}$.  UV continuum and line emission is shock-excited on the stellar surface by material falling from the circum-protostellar disk's inner radius at $R_x = 3.5~R$.  Emission from the postshock material is calculated according to \cite{Pittman+2022}, using the methods of \cite{Calvet+1998} and \cite{Robinson+2019}.  

\begin{figure*}
    \includegraphics[width=0.99\textwidth]{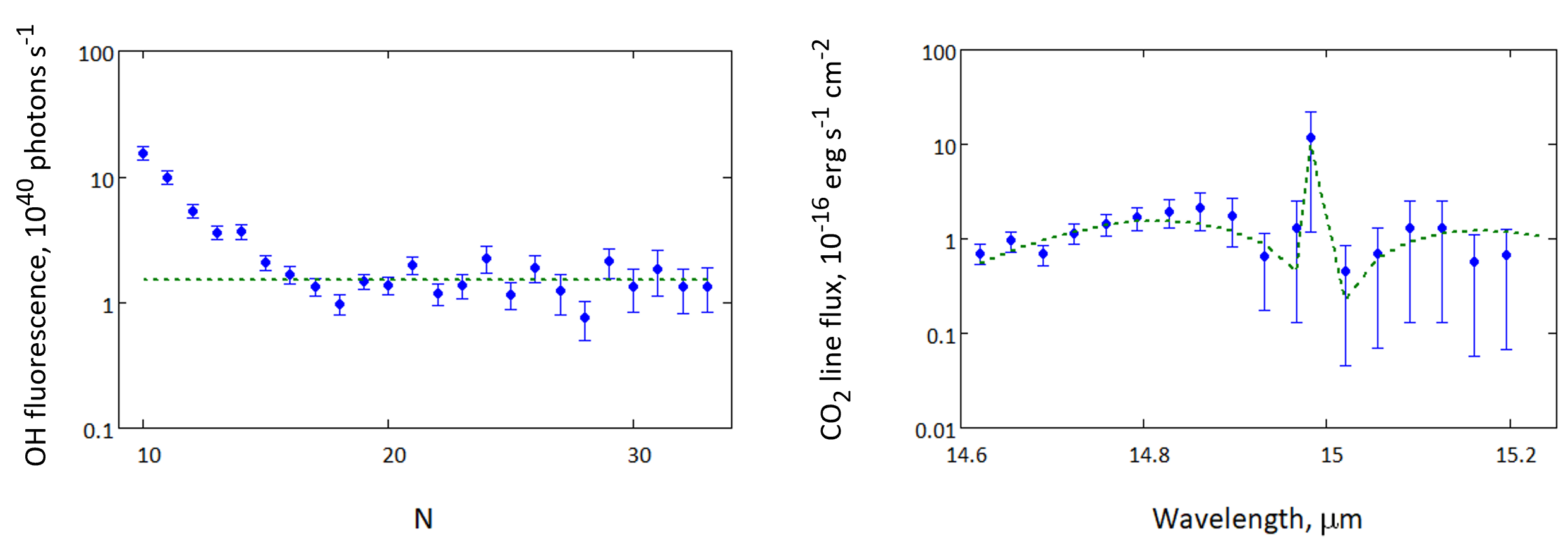}
    \caption{
    Left: extinction-corrected OH-line fluorescence luminosity per rotational transition, $\dot{n}_{\rm f}(N)$. The dashed line indicates the extinction-corrected $\dot{n}_{\rm f}(N)$ averaged over the range $N \geq 17$, from Table \ref{extinction_table}.  
    Right: extinction-corrected $\rm CO_2$-line flux, $f_{\rm CO_2}$. The dashed curve is the fitted emission for $n_{\rm CO_2}$ and $T$ given in Table \ref{extinction_table}. Uncertainties in the plot are dominated by those propagated from the extinction correction, and are particularly large near the center of the $\rm CO_2$ ice feature. 
    \label{extinction_figure}}
\end{figure*}
\begin{deluxetable*}{l l l}
    \tablewidth{0.99\textwidth}
    \tablecaption{Extinction-correction fit to OH and $\rm CO_2$ line fluxes}
    \label{extinction_table}
    \startdata \\
    Extinction column density, $\mathcal{N}_{ext}$ & $(7.3\pm 0.8)\times 10^{22}$ & $\rm H~atoms~cm^{-2}$ \\
    Visual extinction, $A_V$ & $26.7\pm 3.0$ & mag \\
    Fluorescent OH luminosity per rotational transition, $\langle\dot{n}_{\rm f}\rangle$ & $\left( 1.5\pm 0.6\right)\times 10^{40}$ & photons $\rm s^{-1}$ \\
    $\rm CO_2$ gas, $n_{\rm CO_2}$ & $\left( 1.6\pm 0.5\right)\times 10^{44}$ & molecules \\
    $\rm CO_2$ gas temperature, $T$ & $110.0\pm 3.6$ & K \\
    Reduced $\chi^2$ & 1.1 &  \\
    \enddata
\end{deluxetable*}

To this we add observations of line emission not represented in the accretion-shock models, and thought to be produced by gas in the star's magnetopheric accretion funnel \citep{Muzerolle+2001}.  The brightest of these lines, H~I~Ly$\alpha$ $\lambda = 121.6~{\rm nm}$, lies within the SAB.  In a sample of 16 accreting stars, \cite{France+2014} found that Ly~$\alpha$ flux is proportional to, and $10.5\pm 6.6$ times larger than, the integrated $\lambda = 91.2-165~{\rm nm}$ continuum.  We added this continuum-scaled line emission to the accretion models,  resulting in the model spectra shown in Figure \ref{accretion_models}.  The total accretion-related luminosity of the models is $L_{acc} = 3.0\times 10^{-3}~L_\sun\times\left(\dot{M}_a/10^{-9}~M_\sun~{\rm year^{-1}}\right)$.

\subsection{Extinction-corrected parameters of hydroxl and carbon dioxide}
    \label{extinction}
    From our spectra we can obtain extinction-corrected values of the OH fluorescence rate $\dot{n}_{\rm f}$, and the number and temperature of $\rm CO_2$ molecules, $n_{\rm CO_2}$ and $T$.  Our observations offer two independent measurements of reddening. First is the set of OH emission lines, for which each rotational transition in the fluorescent cascade has the same total photon flux.  We calculate wavelength-dependent reddening by ratio of OH photon flux $\dot{n}_{\rm f}(N)$ to its average value over a range of $N$, $\langle{\dot{n}_{\rm f}}\rangle$; that is, departure from a uniform value of ${\dot{n}_{\rm f}}(N)$ can be ascribed to extinction.  To ensure the exclusion of OH emission from mechanisms besides that from the fluorescent cascade, we use only the lines with $N \ge 17$.  This range spans a substantial portion of the broad silicate- and water-ice-related extinction evident in Figure \ref{Spectra}. 

Second is the set of $\rm CO_2$ emission lines, which we assume to arise in the same material as the OH lines.  Like OH, these lines appear to be optically thin: even if the emission source were less than a few AU in radius and the lines were only thermally broadened, they would have peak optical depth $\tau\lesssim 10^{-3}$. We lack accurate excitation rate coefficients for $\rm CO_2 - H_2$ collisions with which to model the populations of $\rm CO_2$'s states, so we assume a local thermal equilibrium (LTE) distribution at temperature $T$.  This leads to an acceptable account of the spectrum, as shown in Figure \ref{extinction_figure}.  We also show in Appendix \ref{water_ii} that the LTE approximation is reasonable, and that it gives the number of $\rm CO_2$ molecules accurately; in other words, that optical pumping \citep{vanGelder+2024a} is negligible compared to collisions in populating the upper states of the detected lines.  

The $\rm CO_2$ flux ratios determine the reddening within the $\rm CO_2$ ice feature.  Combined with a suitable extinction model, either or both of the OH and $\rm CO_2$ measurements leads to a total extinction determination. We include both \footnote{``Either'' instead of ``both'' leads to two closely similar results.} in a $\chi^2$-minimization fitting procedure, using as free parameters the average OH photon flux $\langle{\dot{n}_{\rm f}}\rangle$ in the fluorescent cascade; the temperature $T$ and total number of $\rm CO_2$ molecules $n_{\rm CO_2}$; and the column density $\mathcal{N}_{ext}$ of extinguishing material, assumed to be a nonemitting foreground screen.  As in \cite{Narang+2024}, we adopt the KP5 extinction model \citep{Pontoppidan+2024}.  KP5 has been constructed from optical properties of icy dust grains and constrained by infrared and submillimeter extinction measurements in dense molecular clouds (\citealt{Chapman+2009}, \citealt{Shirley+2011}).  KP5 incorporates a different ice mixture and temperature ($\mathcal{N}({\rm H_2O})/\mathcal{N}({\rm CO_2}) = 5,~T = 20~K$) than we found to be a good match for the I16253 $\rm CO_2$ ice feature. However, the shapes of the two mixtures' ice features \citep{Ehrenfreund+1999} are similar, particularly along their short wavelength edge where the strongest $\rm CO_2$ emission-line constraints lie, in the form of the $\rm CO_2$ R branch. To adapt KP5 to the mixture we infer would make negligible difference in the extinction result.  
Results of the fit and extinction correction appear in Figure \ref{extinction_figure} and Table \ref{extinction_table}.  Our derived value of visual extinction, $26.7\pm 3.0$ mag, agrees well with that derived from $\rm H_2$ emission within our aperture, but associated with I16253's jet or wider-angle outflow: $A_V = 26.5\pm 0.5$ mag \citep{Narang+2024}.  Under the assumption that extinction between central star and molecules is negligible, radiative equilibrium determines a typical stellocentric distance $r = 2.5~{\rm AU}$ for the gas-phase $\rm CO_2$.

\subsection{Mass accretion rate of I16253} 
    \label{I16253}  
    The mass accretion rate $\dot{M}_a$ responsible for the $\rm H_2O$-dissociating UV is determined by $\dot{n}_{\rm f}$ and $\lambda F_\lambda$, calculated above, and the steady-state number of water molecules exposed to protostellar UV, $n_{\rm H_2O}$.  We do not detect this $\rm H_2O$, but it is necessarily located with the emitting OH, and therefore with the $\rm CO_2$.   In the envelope of I16253, where $\rm H_2O$ and $\rm CO_2$ are largely frozen into grains, we found above that $\mathcal{N}(\rm{H_2O})/\mathcal{N}(\rm{CO_2}) = 1.9$ (Section \ref{analysis}) along the sightline toward the central protostar.  This icy dust is representative of the original ingredients of the warmer central structures of I16253.  Thermal and chemical processes could have modified the relative abundances considerably in the inner parts of I16253, but such changes are not currently constrained by observations.  In anticipation of future detection of $\rm H_2O$ emission from I16253, or chemical modelling beyond the scope of this article, we assume, consistent with our observations, that the OH-parental $\rm H_2O$ follows the $\rm CO_2$ at the relative abundance observed in the $\rm CO_2$ ice feature, and we show in Appendices \ref{water_ii} and \ref{water_iii} that this assumption is plausible.  

To obtain $\dot{M}_a$, we consider three scenarios:  one, {\it i}, in which the $\rm H_2O$-dissociating photons exhaust the supply of water; and two others, {\it ii} and {\it iii}, in which the $\rm H_2O$-dissociating photons are outnumbered by water molecules.   The latter scenarios differ in the state of the reservoir of $\rm CO_2$ and $\rm H_2O$: gas phase or solid phase.  

\par\smallskip\noindent 
{\it Scenario i.} The central star's UV photon luminosity in the SAB, capable of dissociating water into OH molecules in the states we see, is $\dot{n}_{\rm SAB} = \frac{4\pi d_0^2}{hc}\int_{SAB}\lambda F_\lambda (\lambda) d\lambda$.  All of our accretion models (section \ref{accretion}) turn out to have $\dot{M}_{\rm model} / \dot{n}_{\rm SAB, model} = 3.2\times 10^{-25}$ gm.  If the observed OH fluorescence rate $\langle\dot{n}_{\rm f}\rangle$ equals or exceeds the rate at which $\rm H_2O$ molecules are provided, then $\dot{n}_{\rm SAB}$ is no smaller than $\langle\dot{n}_{\rm f}\rangle$:

\begin{alignat}{2}
\dot{M}_a (i) &= \langle\dot{n}_{\rm f}\rangle\dot{M}_{\rm model} / \dot{n}_{\rm SAB, model} \nonumber \\ 
&> 4.2\times 10^{-11}~M_\sun~\rm year^{-1}~~~,
\label{Mdot_LL}
\end{alignat}

\noindent at 95\% confidence. 

\par\smallskip\noindent
{\it Scenario ii.} Take the parent $\rm H_2O$ to be gas-phase molecules belonging to one of the fluid components of the central star's surroundings, with its chemical populations in steady state, and lying where the starlight heating of dust matches the temperature determined for $\rm CO_2$ (Section \ref{extinction}).  Possible locations are the innermost outflow-cavity wall, as in a protostellar hot core; in the innermost part of I16253's jet; or interior to the ice line in the embedded disk. In the latter two locations it would intercept a relatively small fraction of the stellar UV, while the hot core would cover most of the central star's sky.  In Appendix \ref{water_ii} we assume such a hot core to exist, and calculate $n_{\rm H_2O}$, obtaining

\begin{equation}
    n_{\rm H_2O}(ii) =  (3.0\pm 1.0)\times 10^{44}~{\rm molecules}~~~.
    \label{water_hot_core}
\end{equation}
\par\smallskip\noindent
For $T = 110~{\rm K}$, none of the $\lambda = 5-27~\mu\rm m$ emission lines of this $\rm H_2 O$ population would lie above the noise in our observations.

\par\smallskip\noindent 
{\it Scenario iii.} Take icy solids --- dust grains or larger bodies --- to be the reservoir of $\rm H_2O$ and $\rm CO_2$.  The molecules would be photodesorbed into the gas phase and then photolyzed, in the manner of solar-system comets and their comae (e.g. \citealt{Crovisier_1989}).  These grains could lie anywhere within UV range of the star, including the outflow cavity or disk atmosphere as well as the cavity walls. In Appendix \ref{water_ii} we correct for the differences in desorption and subsequent photolysis between $\rm H_2O$ and $\rm CO_2$, thereby obtaining 

\begin{equation}
n_{\rm H_2O}(iii) = (5.1\pm 1.9)\times 10^{44}~{\rm molecules  }~~~. 
\label{water_cometary}
\end{equation}

OH in the states we detect is produced at the rate $\dot{n}_{\rm OH}(F_\lambda ) = n_{\rm H_2O}Q\left(\sigma_{SAB}, F_\lambda \right)$, where $Q$ is the photolysis rate of gas-phase $\rm H_2O$ (see Appendix \ref{water_iii}).  $\dot{n}_{\rm OH}$ depends upon the accretion rate $\dot{M}_a$ through $F_\lambda$; $\dot{M}_a$ is determined by equality of $\dot{n}_{\rm OH}$ to the observed OH fluorescence rate $\langle\dot{n}_{\rm f}\rangle$. We  calculated $\dot{n}_{\rm OH}$ for our grid of accretion models, and interpolated among the results to find $\dot{n}_{\rm OH} = \langle\dot{n}_{\rm f}\rangle$.  Thereby our two estimates of $n_{\rm H_2O}$ lead to  

\begin{alignat}{3}
\dot{M}_a (ii) &= \left( 4.1\pm 2.9\right)\times 10^{-10}~M_\sun~\rm year^{-1} \nonumber \\
\dot{M}_a (iii) &= \left( 2.5\pm 1.7\right)\times 10^{-10}~M_\sun~\rm year^{-1} \nonumber \\
\dot{M}_a ({\rm adopted}) &= \left( 3.3\pm 2.2\right)\times 10^{-10}~M_\sun~\rm year^{-1} ~~~,
\label{Mdot}
\end{alignat}

\noindent where the adopted value is the average of the other two, and all are consistent with Equation \ref{Mdot_LL}.  A linear combination of the scenarios {\it ii} and {\it iii} would also be plausible, and would perforce give a result similar to Equation \ref{Mdot}.  All these accretion rate estimates are vastly less than that inferred for I16253 $10^4$ years ago, when its now-0.1-pc-long bipolar molecular outflow began to form: $\dot{M}_a \approx 5\times 10^{-7}~M_\sun~\rm year^{-1}$ \citep{Hsieh+2017}. 
    
\section{Discussion}
    \begin{figure}[htbt]
    \includegraphics[width=0.47\textwidth]{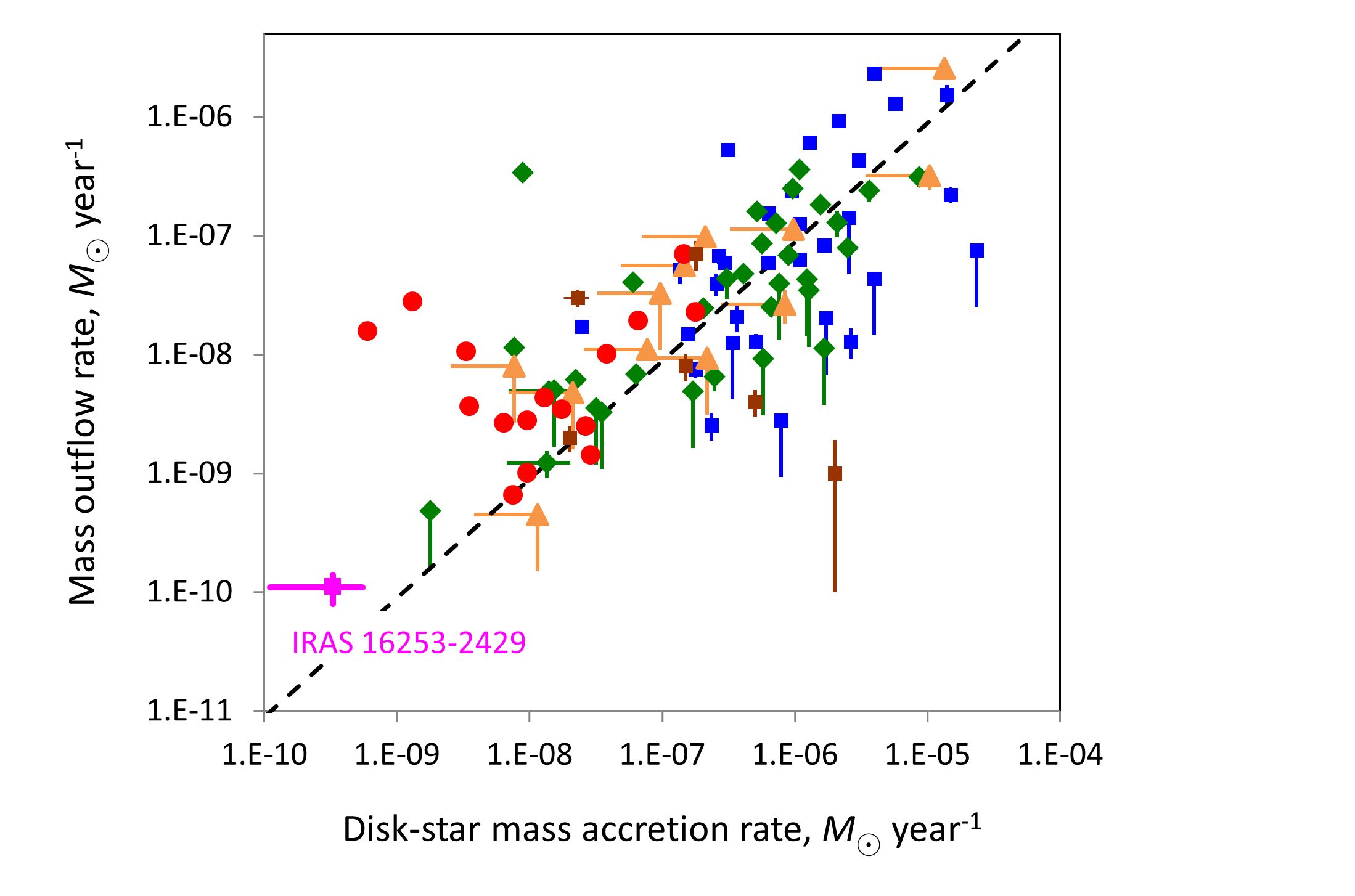}
    \caption{
    Accretion and outflow rates, $\dot{M}_a$ and $\dot{M}_w$, for I16253 (present work, pink square) overlain on accretion and outflow rates from a survey by \cite{Watson+2016} of Class 0/I/flat-spectrum protostars (blue squares, green diamonds, and orange triangles, respectively); of low-mass Class 0 protostars and proto-brown dwarfs (\citealt{Riaz+2021}, brown squares); and of selected Class II YSOs (\citealt{Hartigan+1995}, \citealt{Gullbring+1998}; red circles). The dashed line marks $\dot{M}_w = 0.1\dot{M}_a$.  
    \label{ps_evol_diagram}}
\end{figure}

Equations \ref{Mdot} formally give lower limits to $\dot{M}_a$.  Some of the SAB photons may be absorbed on their way to the $\rm H_2O$, by dust or other molecules.  Neither of the locations posited for the OH-parental population are guaranteed to fill the central star's sky, so UV photons may also escape to larger radii.  We can note, however, that suitable absorbing material is seen to be distributed all around the central star in projection (\citealt{Federman+2023}, \citealt{Federman+2024}, \citealt{Rubinstein+2024}).  Certainly a large fraction of the sky seen from the central star is filled with nearby icy solids which would fit in scenario $\it iii$.  The absorbing material that produces the $\rm CO_2$ ice feature (Section \ref{analysis}) lies far from the disk/envelope midplane as seen from the star. I16253's disk axis is inclined $64.1^\circ$ from the plane of our sky \citep{Aso+2023}; our line of sight has ``declination'' $25.9^\circ$ in I16253's sky.   Accordingly, 44\% of the protostar's sky lies at smaller ``declination'' than we do, along which directions the protostar's UV would encounter material in an even better position to absorb $\rm H_2O$-dissociating UV than that directly along our line of sight.  We conclude that the resulting value of $\dot{M}_a (iii)$ accounts for a large fraction of the accretion-related UV, and that the actual accretion rate cannot be much larger than that given in Equations \ref{Mdot}.

Occupying a much smaller portion of the protostar's sky, I16253's embedded disk intercepts far less of the dissociating UV than would the $\rm H_2 O$  reservoirs of scenarios {\it ii.} and {\it iii.}  A Class II YSO, lacking these additional reservoirs, would not produce OH line emission as prominently as I16253, relative to $\rm H_2 O$ emission.  This agrees with observations (e.g. Figure \ref{Spectra}).  As for the Class II YSOs with exceptional OH emission which we list in Section \ref{fluorescence}: DG Tau has a residual envelope which is massive by Class II standards, evident in extended, scattered visible light \citep{Stapelfeldt+1997} and a flat mid-IR to submillimeter spectral energy distribution \citep{Ohashi+2023}. That some of this material covers large portions of the central star's sky is evident in the starlight's large visible polarization \citep{Bastien_1982}: much of the detected starlight has been scattered, despite the small angle between DG Tau's disk axis and our line of sight \citep{Ohashi+2023}.  Thus DG Tau very probably has a reservoir of $\rm H_2 O$ with distribution similar to I16253's.  We can only speculate that TW Hya's exceptional OH emission is related to the structure of its transitional disk (e.g. \citealt{Menu+2014}). 

In Figure \ref{ps_evol_diagram} we show I16253's mass accretion rate in the context of other, mostly larger-mass or larger-luminosity protostars.  I16253 has a high-speed jet, appearing in the ionic and atomic lines, which dominates mass loss from the central few tens of AU, and from which \cite{Narang+2024} infer mass flow rate $\dot{M}_w = (1.1\pm 0.3)\times 10^{-10} ~M_\sun ~ {\rm year^{-1}}$. This, plus the present result, place I16253 well within the outflow-accretion trend for YSOs, and close to the canonical $\dot{M}_w \sim 0.1\dot{M}_a$ of magnetocentrifugally-accelerated outflows (e.g. \citealt{Pelletier+1992}). As \cite{Narang+2024} show, I16253's jet is highly collimated (opening angle $\lesssim 3^\circ$), requiring that the jet and accretion disk are strongly linked.  Thus the small values of $\dot{M}_a$ and $\dot{M}_w$ do not suggest an unusual arrangement for the footpoints of the magnetic field which accelerates I16253's outflow. 

I16253's luminosity, $L_{bol} = 0.16 ~L_\sun$, is dominated by intrinsic luminosity, with only a 1\% contribution by accretion: $L_{acc} = 1.6\times 10^{-3}~L_\sun$.  By its mass ($M = 0.15~M_\sun$) and intrinsic luminosity, I16253's central star lies close to the stellar mass-luminosity birthline
estimated by \cite{Hartmann+2025}, as do several other Class 0/I systems with measured stellar masses \citep{Hartmann+2025}.  Coupled with the large mass of its infalling envelope, these properties show that I16253 is certainly an early-stage protostar, with an age well within the $\sim 0.5~\rm Myr$ span during which protostars accrete most of their mass (e.g. \citealt{Dunham+2014}, \citealt{Fischer+2017}, \citealt{Federman+2023}).  In I16253 we measure much smaller accretion and outflow rates than those typically reported in protostars, as is clear in Figure \ref{ps_evol_diagram},  and apparently than those of I16253 itself some $10^4$ years ago.  At I16253's current accretion rate it would have taken about $M/\dot{M}_{a} = 450~{\rm Myr}$ to accumulate its current mass (see also \citealt{Narang+2024}). This is qualitatively similar to the finding that several other Class 0/I systems are currently accreting at rates which are not significant over the 0.5 Myr protostellar lifetime \citep{Hartmann+2025}.

Models of protostars which assume the evolution of luminosity to be determined by the evolution of envelope-disk mass infall alone, and decreasing on a $\sim 0.5~\rm Myr$ time scale, take the rate of disk-star mass accretion to be proportional to the rate of envelope-disk infall (e.g. \citealt{Offner+2011}, \citealt{Fischer+2017}).  In such models, the luminosities in the Class 0 phase are dominated by the slowly-evolving steady disk-star accretion, and the dispersion of the observed luminosities at a given mass would be the result of different infall rates for objects with different final masses.  It is otherwise if there is episodic, rapid modulation of disk-star accretion apart from the envelope-disk infall trend, as is canonically invoked for outbursts in YSOs \citep[e.g.][]{Kenyon+1990, 2005ApJ...620L.107M,2009ApJ...692L..67A,2011A&A...526L...1C,Fischer+2012,Safron+2015,2017ApJ...843...45K,2018ApJ...869..146H,2019ApJ...872..183F,2024MNRAS.528.1823C,2024RNAAS...8...64K}. Then the variability due to episodic accretion contributes to, and may even dominate, the luminosity dispersion and mass assembly \citep{Dunham+2010,2012ApJ...747...52D,Zakri+2022,2023ASPC..534..355F}.  Current observations of protostars with measured masses are indeed suggestive of a bimodal distribution of luminosity at a given mass \citep{Hartmann+2025}.  I16253 fits the episodic-accretion picture: its small luminosity compared to other, otherwise similar protostars is due to it being in a phase of very small accretion rate between large accretion-rate episodes.

\section{Conclusions} 
    \begin{enumerate}
\item 
OH pure-rotational emission observed in I16253 is a fluorescent cascade, with the molecules produced in circum-protostellar material by UV dissociation of $\rm H_2O$.  Thereby we have indirectly detected the protostar's accretion-powered UV.  
\item 
Using the reddening of the observed OH and $\rm CO_2$ emission to correct for extinction toward this gas, and the $\rm CO_2$ emission to measure indirectly the (undetected) parent $\rm H_2O$ population, we obtain the accretion rate of I16253: $\dot{M}_a = (3.3\pm 2.2)\times 10^{-10} ~M_\sun ~ {\rm year^{-1}}$.  The primary uncertainty in $\dot{M}_a$ is in estimation of the number of $\rm H_2O$ molecules absorbing the UV. 
\item 
I16253's accretion rate $\dot{M}_a$ is consistent with its small outflow rate $\dot{M}_w$, and with the observational trend centered on the canonical YSO outflow/accretion mass-flow ratio of 0.1.     
\item 
The accretion rate of I16253 is small on the scale of Class 0 protostars, even the other low-mass Class 0 protostars also plotted in Figure \ref{ps_evol_diagram}; it is insufficient by far to form a protostar during a protostellar lifetime.  We therefore have the first clear detection of a Class 0 protostar in a quiescent phase, in which its luminosity is dominated by the intrinsic luminosity of the central star. This discovery contributes to the growing body of evidence that Class 0 protostars undergo episodic accretion with large variation of accretion rates.  
\end{enumerate}
\smallskip

\section{Acknowledgments} 
    We are grateful to Klaus Pontoppidan for providing the KP5 extinction model in advance of publication. We are also grateful to those who maintain the databases ancillary to this work: HITRAN, the NASA Astrophysical Data System, the Leiden Photodissociation and Photoionization database, and the Leiden Ice Database for Astrochemistry. This work is part of JWST general observing program 1802 (Investigating Prostellar Accretion); it is supported in part by NASA \textit{via} the Space Telescope Science Institute.  N.J.E.  thanks the Astronomy Department of the University of Texas for research support.  A.C.G. acknowledges support from PRIN-MUR 2022 20228JPA3A “The path to star and planet formation in the JWST era (PATH)” funded by NextGeneration EU and by INAF-GoG 2022 “NIR-dark Accretion Outbursts in Massive Young stellar objects (NAOMY)” and Large Grant INAF 2022 “YSOs Outflows, Disks and Accretion: towards a global framework for the evolution of planet forming systems (YODA)”. C.P. acknowledges funding from the NSF Graduate Research Fellowship Program under grant No. DGE-1840990.  A.S. gratefully acknowledges support by the Fondecyt Regular (project 
code 1220610), and ANID BASAL project FB210003. Part of this research by MN was carried out at the Jet Propulsion
Laboratory, California Institute of Technology, under a contract with the National Aeronautics and Space Administration (80NM0018D0004).

\facility{JWST (MIRI)}

\section*{Data availability}
{All the data we present in this article were obtained from the Mikulski Archive for Space Telescopes (MAST) at the Space Telescope Science Institute. The specific observations can be accessed via \dataset[DOI: 10.17909/3kky-t040]{https://doi.org/10.17909/3kky-t040}.}
    
\appendix
\section{Mid-infrared spectroscopy of hydroxyl and carbon dioxide}
    \label{Appendix}
    {\bf OH.} OH's unpaired electron, specifically its electron-spin -- electron-orbit interaction (i.e. fine structure), splits the OH ground electronic configuration ($X$) into two components, $X {}^2\Pi_{3/2}$ (X1.5) and $X {}^2\Pi_{1/2}$ (X0.5), as shown in Figure \ref{energy_levels}.  The coupling of OH's angular momentum components conforms well to Hund's Case B (e.g. \citealt{Townes+1975}, chapter 7). Each rotational state within each vibrational manifold of X1.5 and X0.5 is further split by the interaction of the unpaired electron's spin with molecular rotation (i.e. $\Lambda$-type doubling) into two states with opposite total parity but the same total angular momentum quantum number $J$.  Conventionally either the ``rotationless'' parity, labelled $e$ or $f$, or the wavefunction symmetry/antisymmetry about the OH rotational plane, labelled $A'$ and $A''$, is specified for the $\Lambda$-doublet states, as either is conserved in rotational transitions, whereas total parity changes in the transitions. Antisymmetric, $A''$ states lie higher in energy than the symmetric, $A'$ states.  For the range of concern here, $f$ states lie higher in energy than $e$ for X1.5, and \textit{vice versa} for X0.5.  The molecular total orbital angular momentum $N\hbar$ --- including rotation, but not electron spin --- is approximately conserved, with its index $N = J\pm 1/2$ serving as a good quantum number.  The selection rules for OH's electric dipole transitions are thus  $\Delta N = \pm 1, 0$ but no $0\leftrightarrow 0$, and no $A''\rightleftharpoons A'$.  Each transition $\Delta N$ corresponds to a quartet of OH lines, as in Figures \ref{Spectra} and \ref{energy_levels}.

$\bf CO_2.$ See \cite{Townes+1975}, chapter 2, for nomenclature. Each of the ro-vibration lines of $\rm CO_2$ is split into two components by $\ell$-type doubling, which are spectrally unresolved in our observations.  The symmetry of $\rm CO_2$, and the zero spin of the oxygen nuclei, forbid odd values of the total angular momentum quantum number $J$ in the ground vibrational state, $\nu_1 \nu_2^\ell \nu_3 = 0 0^0 0$. 

\section{The fraction of UV photons absorbed by water} 
    \subsection{Scenario ii: carbon dioxide and water in the hot core of I16253}
    \label{water_ii}  
    Here we take the observed gas-phase $\rm CO_2$ to lie along the innermost parts of I16253's outflow-cavity wall, as in a protostellar hot core.  In Section \ref{extinction}, we take the rotation-vibration states of the $\rm CO_2$ lines to be populated thermally, and calculate the extinction-corrected number and LTE temperature of the molecules (Table \ref{extinction_table}), whence the number of water molecules in the population is simply 

\begin{equation}
    n_{\rm H_2O}(ii) = \frac{\mathcal{N}(\rm{H_2O})}{\mathcal{N}(\rm{CO_2})}n_{\rm CO_2} 
    = (3.0\pm 1.0)\times 10^{44}~{\rm molecules}~~~.
    \label{simplest}
\end{equation}

LTE is not an unreasonable initial assumption in light of the large mass of I16253's envelope ($1 \ M_\sun$), the consequent large density of the protostellar core, and current incompleteness in molecular-physics data which would be needed in a more detailed model. Assuming radiative equilibrium for dust grains, and gas heated by collisions with dust, the $T = 110$ K molecules lie approximately $r_{in} = 2.5~\rm{AU}$ from the central star.  Models of Class 0 envelopes by \cite{Kristensen+2012} typically have power-law density profile with index $p = 1.5$, and outer-to-inner radius ratio $Y = r_{out}/r_{in} = 1000$. Application of these parameters to I16253 gives hydrogen number density $\sim 5 \times 10^{10}~{\rm cm}^{-3}$ at $r = r_{in}$ --- indeed large on the scale of hot cores --- and containin $n_{\rm H_2} \sim 7 \times 10^{52}$ hydrogen molecules in the innermost warm zone in which most ices are completely sublimated ($T \geq~90~{\rm K}$).      

LTE also assumes that infrared pumping does not substantially affect the populations of the $\rm CO_2$ rotation-vibration states, as it does in many hot cores (e.g. \citealt{vanGelder+2024b}). We can check this by considering the continuum brightness temperature $T_b$ of I16253 at the $\rm CO_2$ lines' frequencies ($\lambda \sim 15~\mu {\rm m}, \nu \sim 2\times 10^{13}~{\rm Hz}$). Correcting the observed flux density (1.8 mJy) for extinction toward and distance to I16253's central few AU to give the continuum surface brightness $I_\nu$, we get

\begin{equation}
T_b = \frac{h\nu}{k}\frac{1}{{\rm ln}\left(1 + \frac{2h\nu^3}{c^2 I_\nu}\right)} = 45~{\rm K}~~~,
\label{T_b}
\end{equation}

\noindent small enough compared to the LTE excitation temperature of $T = 110~{\rm K}$ to conclude that infrared pumping can be neglected in first approximation. 

Observations (e.g. \citealt{vanDishoeck+2021}, \citealt{vanGelder+2024a}) and models (e.g. \citealt{Notsu+2021}) typically give $\rm H_2 O/H_2 \sim 10^{-7} - 10^{-5}$ in the inner parts of hot cores.  The models also show that $\rm H_2 O/H_2$ and $\rm CO_2 /H_2$ are suppressed in hot cores relative to elsewhere in the protostars' envelopes, which is understood as due to X-ray dissociation of these molecules. A rough estimate based on Equation \ref{simplest}, and the envelope/innermost-warm-zone parameters of I16253, compare well to these results: we obtain $\rm H_2 O/H_2 \sim 4 \times 10^{-7}$.  The hot-core models by Notsu et al. are optimized for protostars significantly more massive than I16253, and there have been no X-ray observations of I16253, so the models may not apply accurately here.  However, it is not unusual in the model results for the interior $\rm H_2 O$/$\rm CO_2$ relative abundance to resemble that of the colder outer parts of the envelope (\citealt{Notsu+2021}, figures 3 and C.1). It is therefore not unreasonable to use the $\rm CO_2$ ice feature's $\rm H_2 O$/$\rm CO_2$ relative abundance in Equation \ref{simplest}.  

Accurate calculations of collisional excitation rate coefficients for the rotation-vibration states of $\rm CO_2$ are forthcoming \citep{Selim+2023}, which will allow precise modeling of the molecular abundances in the core of I16253 in future articles.

    \subsection{Scenario iii: carbon dioxide and water photodesorbed from icy solids near the center of I16253}
    \label{water_iii}  
    
Next we consider the emitting medium to consist of gas recently released from icy solids, and calculate $n_{\rm H_2O}$ from the observed $\rm CO_2$ gas and ice, corrected for the rates of photoproduction and photolysis of $\rm H_2O$ and $\rm CO_2$. Again we neglect extinction between the central star and the icy solids associated with $\rm CO_2$ and OH emission.  

According to the $\rm CO_2$ ice feature shape (Figure \ref{ice_spectrum}), $\rm H_2O$ and $\rm CO_2$ are present in I16253's icy solids in the ratio $\mathcal{N}(\rm{H_2O})/\mathcal{N}(\rm{CO_2}) = 1.9$.  However, the two gas species are not released from the solids at similar rates, nor do they have the same photolysis lifetime, so we need to make two corrections in order to calculate the steady-state number of water molecules $n_{\rm H_2O}$ from $n_{\rm CO_2}$. At the temperature obtained in the fit shown in Figure \ref{ice_spectrum}, $\rm H_2O$ and $\rm CO_2$ ice would sublimate or thermally desorb at negligibly small rates \citep{Minissale+2022}.   But both species are readily photodesorbed by far-UV light. The desorption probability per UV photon, $p(\lambda )$, is calculated for $\rm H_2O$ by \cite{Andersson+2008}, and measured for $\rm CO_2$ by \cite{Fillion+2014}, whence a species' photodesorption rate $q_{mol}$ is determined from our accretion-model flux $\lambda F_\lambda$ at system distance $d_0$, for the solids' stellocentric distance $r$:

\begin{equation}
    q_{mol}\left(p , F_\lambda \right) = {\frac{1}{hc}}{\frac{d_0^2}{r^2}}\int \lambda F_\lambda (\lambda ) p(\lambda ) d\lambda \label{eqn2}
\end{equation}

\noindent The integral is over all wavelengths.  Ratios of $q_{mol}$ turn out to be independent of accretion rate for our model grid. 

Photodissociation and photoionization cross sections are given for $\rm H_2O$, OH and $\rm CO_2$ by \cite{Fillion+2003}, \cite{vanDishoeck+1984}, and \cite{Huestis+2010} respectively.  \cite{Heays+2017} parse the photodissociation cross section of $\rm H_2O$ into components with specific products, including OH($X^2\Pi$) + H \footnote{\label{leiden_pd} See  https://home.strw.leidenuniv.nl/$\sim$ewine/photo/.}.  In turn this latter cross-section component separates cleanly into the FAB and SAB, yielding the cross section $\sigma_{SAB}$ which gives rise to OH rotational fluorescence.   
  
In analogy with Equation \ref{eqn2}, a species' total photolysis rate per molecule, $Q_{mol}$, is

\begin{equation}
    Q_{mol}\left(\sigma_{t} , F_\lambda \right) = {\frac{1}{hc}}{\frac{d_0^2}{r^2}}\int \lambda F_\lambda (\lambda )\sigma_{t} (\lambda ) d\lambda~~~, 
    \label{eqn1}
\end{equation}

\noindent where  $\sigma_{t} (\lambda )$ the sum of the molecule's photodissociation and photoionization cross sections.  Using the $\dot{M}_a = 10^{-9} ~M_\sun~{\rm year^{-1}}$ accretion model for example, we obtain photolysis lifetimes $1/Q_{\rm H_2O} = 5.4\times 10^3$~sec, $1/Q_{OH} = 1.7\times 10^4$~sec, and $1/Q_{\rm CO_2} = 2.3\times 10^5$~sec for $\rm H_2O$, OH, and $\rm CO_2$ respectively. Thus, under conditions likely to prevail in I16253, $\rm H_2O$ is most rapidly destroyed of the three species; OH decays in its fluorescent cascade to small values of $N$ long before being destroyed; and $\rm CO_2$, outlasting both, subsists long enough to serve as a proxy for the parent $\rm H_2O$.   As also is true for $q_{mol}$, ratios of $Q_{mol}$ turn out to be independent of accretion rate for our model grid. 

Thus the steady-state number of molecules in the parent $\rm H_2O$ population, within the photolysis lifetime after release from the ice, is

\begin{equation}
n_{\rm H_2O}(iii)
= 
\frac{{\mathcal{N}({\rm H_2O})}}{{\mathcal{N}({\rm CO_2})}} \frac{q_{\rm H_2O}}{q_{\rm CO_2}} \frac{Q_{\rm CO_2}}{Q_{\rm H_2O}} n_{\rm CO_2} 
= (1.9)(71.2)(0.0237)n_{\rm CO_2} 
= (5.1\pm 1.9)\times 10^{44}~~~,  
\end{equation}

\noindent within uncertainties of $n_{\rm H_2O}(ii)$.  Note that the dependence on stellocentric distance $r$ has canceled out, as usual for processes initiated by single photons.  The irradiated and desorbed dust grains could lie anywhere SAB UV can reach within the unresolved source ($r < 40 \rm AU$).  In this case the temperature $T$ determined from the fit to the spectral-line fluxes would be characteristic of the kinetic energy $\Delta E$ of the $\rm CO_2$ molecules upon UV-induced release from the dust grains, for which $\Delta E/k = 110~\rm{K}$ would not be unreasonable.

\clearpage
\startlongtable 
\begin{deluxetable*}{llllllrllllll}
    \tablewidth{0pt}
    \tablecaption{$\lambda~=~10-26~\micron$ spectral line fluxes within 1 arcsec of I16253\label{list_of_fluxes}}
    \tablehead{
    \multicolumn{1}{c}{Wavelength} & \multicolumn{4}{c}{Species and transition}  & & \multicolumn{4}{c}{Observed flux,} & \multicolumn{3}{c}{Extinction}  \\
     \multicolumn{1}{c}{$\micron$} & \multicolumn{4}{c}{(see endnotes for description)}  & & \multicolumn{4}{c}{$\rm{10^{-17}~erg~sec^{-1}~cm^{-2}}$}   &  \multicolumn{3}{c}{factor} \\
      & & & & & & & \\
    \multicolumn{1}{l}{\tablenotemark{\rm 1}} &   \multicolumn{1}{c}{\tablenotemark{\rm 2}} &  \multicolumn{1}{c}{\tablenotemark{\rm 3}} &  \multicolumn{1}{l}{\tablenotemark{\rm 4}} &  \multicolumn{1}{l}{\tablenotemark{\rm 5}} &  \multicolumn{1}{l}{\tablenotemark{\rm 6}} &  \multicolumn{4}{c}{\tablenotemark{\rm 7}} &  \multicolumn{3}{c}{\tablenotemark{\rm 8}}  \\
    }
    \decimals
    \startdata
10.060	&	OH	&	X1.5	&	33	&	$A''$	&		&	$<$	&	0.2	&		&		&	49.6	&	$\pm$	&	21.9	\\
10.062	&	OH	&	X0.5	&	33	&	$A''$	&		&	$<$	&	0.2	&		&		&	49.6	&	$\pm$	&	21.9	\\
10.069	&	OH	&	X1.5	&	33	&	$A'$	&	$+$	&		&	1.9	&	$\pm$	&	0.1	&	49.5	&	$\pm$	&	21.8	\\
10.071	&	OH	&	X0.5	&	33	&	$A'$	&		&		&		&		&		&		&		&		\\
10.221	&	OH	&	X1.5	&	32	&	$A''$	&		&	$<$	&	0.2	&		&		&	46.9	&	$\pm$	&	20.4	\\
10.224	&	OH	&	X0.5	&	32	&	$A''$	&		&	$<$	&	0.1	&		&		&	46.8	&	$\pm$	&	20.3	\\
10.231	&	OH	&	X1.5	&	32	&	$A'$	&	$+$	&		&	2.05	&	$\pm$	&	0.04	&	46.8	&	$\pm$	&	20.3	\\
10.233	&	OH	&	X0.5	&	32	&	$A'$	&		&		&		&		&		&		&		&		\\
10.399	&	OH	&	X1.5	&	31	&	$A''$	&		&	$<$	&	0.2	&		&		&	43.2	&	$\pm$	&	18.4	\\
10.402	&	OH	&	X0.5	&	31	&	$A''$	&		&	$<$	&	0.2	&		&		&	43.1	&	$\pm$	&	18.3	\\
10.410	&	OH	&	X1.5	&	31	&	$A'$	&	$+$	&		&	3.20	&	$\pm$	&	0.05	&	42.8	&	$\pm$	&	18.2	\\
10.412	&	OH	&	X0.5	&	31	&	$A'$	&		&		&		&		&		&		&		&		\\
10.595	&	OH	&	X1.5	&	30	&	$A''$	&		&	$<$	&	0.2	&		&		&	37.6	&	$\pm$	&	15.4	\\
10.598	&	OH	&	X0.5	&	30	&	$A''$	&		&	$<$	&	0.2	&		&		&	37.4	&	$\pm$	&	15.3	\\
10.607	&	OH	&	X1.5	&	30	&	$A'$	&	$+$	&		&	2.53	&	$\pm$	&	0.04	&	37.2	&	$\pm$	&	15.2	\\
10.609	&	OH	&	X0.5	&	30	&	$A'$	&		&		&		&		&		&		&		&		\\
10.681	&	[Ni II]	&	${}^4 F_{7/2}-{}^4 F_{9/2}$	&		&		&		&		&	6.07	&	$\pm$	&	0.03	&	35.5	&	$\pm$	&	14.3	\\
10.810	&	OH	&	X1.5	&	29	&	$A''$	&		&	$<$	&	0.1	&		&		&	32.7	&	$\pm$	&	12.9	\\
10.815	&	OH	&	X0.5	&	29	&	$A''$	&		&	$<$	&	0.1	&		&		&	32.6	&	$\pm$	&	12.8	\\
10.823	&	OH	&	X1.5	&	29	&	$A'$	&		&		&	2.7	&	$\pm$	&	0.1	&	32.5	&	$\pm$	&	12.8	\\
10.826	&	OH	&	X0.5	&	29	&	$A'$	&		&		&	2.1	&	$\pm$	&	0.1	&	32.4	&	$\pm$	&	12.8	\\
11.047	&	OH	&	X1.5	&	28	&	$A''$	&		&	$<$	&	0.2	&		&		&	29.7	&	$\pm$	&	11.4	\\
11.052	&	OH	&	X0.5	&	28	&	$A''$	&		&	$<$	&	0.1	&		&		&	29.7	&	$\pm$	&	11.4	\\
11.061	&	OH	&	X1.5	&	28	&	$A'$	&	$+$	&		&	1.65	&	$\pm$	&	0.04	&	29.7	&	$\pm$	&	11.4	\\
11.065	&	OH	&	X0.5	&	28	&	$A'$	&		&		&		&		&		&		&		&		\\
11.307	&	[Ni I]	&	${}^3 F_{2}-{}^3 F_{3}$	&		&		&		&	$<$	&	0.1	&		&		&	27.2	&	$\pm$	&	10.2	\\
11.308	&	OH	&	X1.5	&	27	&	$A''$	&		&	$<$	&	0.1	&		&		&	27.2	&	$\pm$	&	10.2	\\
11.313	&	OH	&	X0.5	&	27	&	$A''$	&		&	$<$	&	0.1	&		&		&	27.1	&	$\pm$	&	10.1	\\
11.323	&	OH	&	X1.5	&	27	&	$A'$	&	$+$	&		&	3.1	&	$\pm$	&	0.1	&	27.0	&	$\pm$	&	10.1	\\
11.327	&	OH	&	X0.5	&	27	&	$A'$	&		&		&		&		&		&		&		&		\\
11.333	&	[Cl I]	&	${}^2 P_{1/2} - {}^2 P_{3/2}$	&		&		&		&		&	3.2	&	$\pm$	&	0.1	&	26.7	&	$\pm$	&	9.9	\\
11.594	&	OH	&	X1.5	&	26	&	$A''$	&		&	$<$	&	0.2	&		&		&	25.1	&	$\pm$	&	9.1	\\
11.600	&	OH	&	X0.5	&	26	&	$A''$	&		&	$<$	&	0.2	&		&		&	24.9	&	$\pm$	&	9.1	\\
11.610	&	OH	&	X1.5	&	26	&	$A'$	&		&		&	2.5	&	$\pm$	&	0.1	&	24.8	&	$\pm$	&	9.0	\\
11.615	&	OH	&	X0.5	&	26	&	$A'$	&		&		&	2.7	&	$\pm$	&	0.1	&	24.8	&	$\pm$	&	9.0	\\
11.909	&	OH	&	X1.5	&	25	&	$A''$	&		&	$<$	&	0.1	&		&		&	22.1	&	$\pm$	&	7.7	\\
11.916	&	OH	&	X0.5	&	25	&	$A''$	&		&	$<$	&	0.1	&		&		&	22.1	&	$\pm$	&	7.7	\\
11.926	&	OH	&	X1.5	&	25	&	$A'$	&		&		&	1.54	&	$\pm$	&	0.02	&	22.1	&	$\pm$	&	7.7	\\
11.931	&	OH	&	X0.5	&	25	&	$A'$	&		&		&	2.05	&	$\pm$	&	0.02	&	22.1	&	$\pm$	&	7.7	\\
12.256	&	OH	&	X1.5	&	24	&	$A''$	&		&		&	0.53	&	$\pm$	&	0.02	&	18.8	&	$\pm$	&	6.2	\\
12.264	&	OH	&	X0.5	&	24	&	$A''$	&		&		&	0.85	&	$\pm$	&	0.02	&	18.7	&	$\pm$	&	6.2	\\
12.274	&	OH	&	X1.5	&	24	&	$A'$	&		&		&	2.7	&	$\pm$	&	0.5	&	18.6	&	$\pm$	&	6.1	\\
12.279	&	$\rm H_2$	&	$v = 0$	&	S 2	&		&		&		&	13.19	&	$\pm$	&	0.02	&	18.5	&	$\pm$	&	6.1	\\
12.280	&	OH	&	X0.5	&	24	&	$A'$	&		&		&	4.3	&	$\pm$	&	0.9	&	18.5	&	$\pm$	&	6.1	\\
12.372	&	H I	&	Hu $\alpha$	&		&		&		&	$<$	&	0.1	&		&		&	17.8	&	$\pm$	&	5.8	\\
12.638	&	OH	&	X1.5	&	23	&	$A''$	&		&	$<$	&	0.1	&		&		&	15.6	&	$\pm$	&	4.8	\\
12.647	&	OH	&	X0.5	&	23	&	$A''$	&		&	$<$	&	0.1	&		&		&	15.5	&	$\pm$	&	4.8	\\
12.657	&	OH	&	X1.5	&	23	&	$A'$	&		&		&	2.93	&	$\pm$	&	0.02	&	15.4	&	$\pm$	&	4.8	\\
12.665	&	OH	&	X0.5	&	23	&	$A'$	&		&		&	2.84	&	$\pm$	&	0.02	&	15.4	&	$\pm$	&	4.8	\\
12.728	&	[Ni II]	&	${}^4 F_{5/2}-{}^4 F_{7/2}$	&		&		&		&		&	2.24	&	$\pm$	&	0.02	&	14.9	&	$\pm$	&	4.5	\\
12.812	&	[Ne II]	&	${}^2 P_{1/2} - {}^2 P_{3/2}$	&		&		&		&		&	29.21	&	$\pm$	&	0.02	&	14.2	&	$\pm$	&	4.3	\\
13.060	&	OH	&	X1.5	&	22	&	$A''$	&		&	$<$	&	0.1	&		&		&	12.7	&	$\pm$	&	3.7	\\
13.071	&	OH	&	X0.5	&	22	&	$A''$	&		&		&	0.40	&	$\pm$	&	0.02	&	12.6	&	$\pm$	&	3.6	\\
13.081	&	OH	&	X1.5	&	22	&	$A'$	&		&		&	3.05	&	$\pm$	&	0.02	&	12.5	&	$\pm$	&	3.6	\\
13.089	&	OH	&	X0.5	&	22	&	$A'$	&		&		&	2.59	&	$\pm$	&	0.02	&	12.4	&	$\pm$	&	3.5	\\
13.527	&	OH	&	X1.5	&	21	&	$A''$	&		&		&	0.68	&	$\pm$	&	0.04	&	10.3	&	$\pm$	&	2.7	\\
13.540	&	OH	&	X0.5	&	21	&	$A''$	&		&		&	1.16	&	$\pm$	&	0.04	&	10.2	&	$\pm$	&	2.7	\\
13.549	&	OH	&	X1.5	&	21	&	$A'$	&		&		&	5.15	&	$\pm$	&	0.04	&	10.2	&	$\pm$	&	2.7	\\
13.560	&	OH	&	X0.5	&	21	&	$A'$	&		&		&	5.14	&	$\pm$	&	0.04	&	10.2	&	$\pm$	&	2.7	\\
14.045	&	OH	&	X1.5	&	20	&	$A''$	&		&	$<$	&	0.5	&		&		&	8.7	&	$\pm$	&	2.1	\\
14.061	&	OH	&	X0.5	&	20	&	$A''$	&		&		&	0.92	&	$\pm$	&	0.05	&	8.7	&	$\pm$	&	2.1	\\
14.069	&	OH	&	X1.5	&	20	&	$A'$	&		&		&	4.0	&	$\pm$	&	0.1	&	8.7	&	$\pm$	&	2.1	\\
14.082	&	OH	&	X0.5	&	20	&	$A'$	&		&		&	4.22	&	$\pm$	&	0.03	&	8.6	&	$\pm$	&	2.1	\\
14.621	&	$\rm CO_2$	&	$0 1^1 0 - 0 0^0 0$	&	R 20	&		&		&		&	0.84	&	$\pm$	&	0.03	&	8.4	&	$\pm$	&	2.0	\\
14.622	&	OH	&	X1.5	&	19	&	$A''$	&		&		&	0.92	&	$\pm$	&	0.03	&	8.4	&	$\pm$	&	2.0	\\
14.641	&	OH	&	X0.5	&	19	&	$A''$	&		&		&	0.7	&	$\pm$	&	0.1	&	8.4	&	$\pm$	&	2.0	\\
14.648	&	OH	&	X1.5	&	19	&	$A'$	&		&		&	4.26	&	$\pm$	&	0.03	&	8.4	&	$\pm$	&	2.0	\\
14.655	&	$\rm CO_2$	&	$0 1^1 0 - 0 0^0 0$	&	R 18	&		&		&		&	1.16	&	$\pm$	&	0.03	&	8.3	&	$\pm$	&	2.0	\\
14.664	&	OH	&	X0.5	&	19	&	$A'$	&		&		&	4.31	&	$\pm$	&	0.03	&	8.3	&	$\pm$	&	2.0	\\
14.689	&	$\rm CO_2$	&	$0 1^1 0 - 0 0^0 0$	&	R 16	&		&		&		&	0.82	&	$\pm$	&	0.03	&	8.4	&	$\pm$	&	2.0	\\
14.724	&	$\rm CO_2$	&	$0 1^1 0 - 0 0^0 0$	&	R 14	&		&		&		&	1.34	&	$\pm$	&	0.03	&	8.6	&	$\pm$	&	2.1	\\
14.758	&	$\rm CO_2$	&	$0 1^1 0 - 0 0^0 0$	&	R 12	&		&		&		&	1.56	&	$\pm$	&	0.03	&	9.1	&	$\pm$	&	2.3	\\
14.792	&	$\rm CO_2$	&	$0 1^1 0 - 0 0^0 0$	&	R 10	&		&		&		&	1.75	&	$\pm$	&	0.03	&	9.6	&	$\pm$	&	2.5	\\
14.827	&	$\rm CO_2$	&	$0 1^1 0 - 0 0^0 0$	&	R 8	&		&		&		&	1.87	&	$\pm$	&	0.03	&	10.4	&	$\pm$	&	2.7	\\
14.862	&	$\rm CO_2$	&	$0 1^1 0 - 0 0^0 0$	&	R 6	&		&		&		&	1.97	&	$\pm$	&	0.03	&	10.8	&	$\pm$	&	2.9	\\
14.897	&	$\rm CO_2$	&	$0 1^1 0 - 0 0^0 0$	&	R 4	&		&		&		&	1.54	&	$\pm$	&	0.03	&	11.4	&	$\pm$	&	3.2	\\
14.931	&	$\rm CO_2$	&	$0 1^1 0 - 0 0^0 0$	&	R 2	&		&		&		&	0.55	&	$\pm$	&	0.03	&	11.8	&	$\pm$	&	3.3	\\
14.966	&	$\rm CO_2$	&	$0 1^1 0 - 0 0^0 0$	&	R 0	&		&		&		&	1.06	&	$\pm$	&	0.03	&	12.3	&	$\pm$	&	3.5	\\
14.982	&	$\rm CO_2$	&	$0 1^1 0 - 0 0^0 0$	&	Q branch	&		&		&		&	9.37	&	$\pm$	&	0.05	&	12.5	&	$\pm$	&	3.6	\\
15.019	&	$\rm CO_2$	&	$0 1^1 0 - 0 0^0 0$	&	P 2	&		&		&		&	0.34	&	$\pm$	&	0.03	&	13.1	&	$\pm$	&	3.8	\\
15.054	&	$\rm CO_2$	&	$0 1^1 0 - 0 0^0 0$	&	P 4	&		&		&		&	0.51	&	$\pm$	&	0.03	&	13.6	&	$\pm$	&	4.0	\\
15.090	&	$\rm CO_2$	&	$0 1^1 0 - 0 0^0 0$	&	P 6	&		&		&		&	0.92	&	$\pm$	&	0.03	&	14.1	&	$\pm$	&	4.2	\\
15.125	&	$\rm CO_2$	&	$0 1^1 0 - 0 0^0 0$	&	P 8	&		&		&		&	0.93	&	$\pm$	&	0.03	&	14.3	&	$\pm$	&	4.3	\\
15.160	&	$\rm CO_2$	&	$0 1^1 0 - 0 0^0 0$	&	P 10	&		&		&		&	0.40	&	$\pm$	&	0.03	&	14.3	&	$\pm$	&	4.3	\\
15.196	&	$\rm CO_2$	&	$0 1^1 0 - 0 0^0 0$	&	P 12	&		&		&		&	0.48	&	$\pm$	&	0.03	&	14.1	&	$\pm$	&	4.2	\\
15.232	&	$\rm CO_2$	&	$0 1^1 0 - 0 0^0 0$	&	P 14	&		&		&		&	0.11	&	$\pm$	&	0.03	&	13.8	&	$\pm$	&	4.1	\\
15.267	&	OH	&	X1.5	&	18	&	$A''$	&		&		&	0.34	&	$\pm$	&	0.03	&	13.5	&	$\pm$	&	4.0	\\
15.291	&	OH	&	X0.5	&	18	&	$A''$	&	$+$	&		&	2.05	&	$\pm$	&	0.03	&	13.0	&	$\pm$	&	3.8	\\
15.295	&	OH	&	X1.5	&	18	&	$A'$	&		&		&		&		&		&		&		&		\\
15.314	&	OH	&	X0.5	&	18	&	$A'$	&		&		&	1.82	&	$\pm$	&	0.03	&	12.6	&	$\pm$	&	3.6	\\
15.991	&	OH	&	X1.5	&	17	&	$A''$	&		&		&	0.85	&	$\pm$	&	0.03	&	8.4	&	$\pm$	&	2.0	\\
16.020	&	OH	&	X0.5	&	17	&	$A''$	&	$+$	&		&	4.36	&	$\pm$	&	0.03	&	8.4	&	$\pm$	&	2.0	\\
16.021	&	OH	&	X1.5	&	17	&	$A'$	&		&		&		&		&		&	8.4	&	$\pm$	&	2.0	\\
16.045	&	OH	&	X0.5	&	17	&	$A'$	&		&		&	3.12	&	$\pm$	&	0.03	&	8.4	&	$\pm$	&	2.0	\\
16.808	&	OH	&	X1.5	&	16	&	$A''$	&		&		&	1.6	&	$\pm$	&	0.03	&	8.7	&	$\pm$	&	2.1	\\
16.840	&	OH	&	X1.5	&	16	&	$A'$	&	$+$	&		&	4.8	&	$\pm$	&	0.03	&	8.8	&	$\pm$	&	2.2	\\
16.843	&	OH	&	X0.5	&	16	&	$A''$	&		&		&		&		&		&		&		&		\\
16.870	&	OH	&	X0.5	&	16	&	$A'$	&		&		&	3.20	&	$\pm$	&	0.03	&	8.8	&	$\pm$	&	2.2	\\
17.035	&	$\rm H_2$	&	$v = 0$	&	S 1	&		&		&		&	7.52	&	$\pm$	&	0.03	&	8.9	&	$\pm$	&	2.2	\\
17.734	&	OH	&	X1.5	&	15	&	$A''$	&		&		&	1.8	&	$\pm$	&	0.03	&	9.2	&	$\pm$	&	2.3	\\
17.769	&	OH	&	X1.5	&	15	&	$A'$	&		&		&	4.1	&	$\pm$	&	0.03	&	9.3	&	$\pm$	&	2.3	\\
17.779	&	OH	&	X0.5	&	15	&	$A''$	&		&		&	0.82	&	$\pm$	&	0.03	&	9.3	&	$\pm$	&	2.3	\\
17.808	&	OH	&	X0.5	&	15	&	$A'$	&		&		&	4.07	&	$\pm$	&	0.03	&	9.3	&	$\pm$	&	2.3	\\
17.934	&	[Fe II]	&	${}^4 F_{7/2}-{}^4 F_{9/2}$	&		&		&		&		&	173.90	&	$\pm$	&	0.02	&	9.2	&	$\pm$	&	2.3	\\
18.792	&	OH	&	X1.5	&	14	&	$A''$	&		&		&	2.5	&	$\pm$	&	0.1	&	9.1	&	$\pm$	&	2.3	\\
18.830	&	OH	&	X1.5	&	14	&	$A'$	&		&		&	6.7	&	$\pm$	&	0.1	&	9.1	&	$\pm$	&	2.3	\\
18.849	&	OH	&	X0.5	&	14	&	$A''$	&		&		&	3.2	&	$\pm$	&	0.1	&	9.1	&	$\pm$	&	2.3	\\
18.880	&	OH	&	X0.5	&	14	&	$A'$	&		&		&	5.7	&	$\pm$	&	0.1	&	9.1	&	$\pm$	&	2.3	\\
20.009	&	OH	&	X1.5	&	13	&	$A''$	&		&		&	2.4	&	$\pm$	&	0.3	&	8.3	&	$\pm$	&	2.0	\\
20.050	&	OH	&	X1.5	&	13	&	$A'$	&		&		&	5.4	&	$\pm$	&	0.3	&	8.3	&	$\pm$	&	2.0	\\
20.082	&	OH	&	X0.5	&	13	&	$A''$	&		&		&	3.3	&	$\pm$	&	0.3	&	8.2	&	$\pm$	&	2.0	\\
20.115	&	OH	&	X0.5	&	13	&	$A'$	&		&		&	7.4	&	$\pm$	&	0.3	&	8.2	&	$\pm$	&	1.9	\\
21.420	&	OH	&	X1.5	&	12	&	$A''$	&		&		&	5.2	&	$\pm$	&	0.5	&	7.1	&	$\pm$	&	1.6	\\
21.465	&	OH	&	X1.5	&	12	&	$A'$	&		&		&	11.7	&	$\pm$	&	0.5	&	7.1	&	$\pm$	&	1.6	\\
21.517	&	OH	&	X0.5	&	12	&	$A''$	&		&		&	3.9	&	$\pm$	&	0.5	&	7.0	&	$\pm$	&	1.6	\\
21.553	&	OH	&	X0.5	&	12	&	$A'$	&		&		&	8.9	&	$\pm$	&	0.5	&	7.0	&	$\pm$	&	1.5	\\
23.074	&	OH	&	X1.5	&	11	&	$A''$	&		&		&	7.6	&	$\pm$	&	0.8	&	6.1	&	$\pm$	&	1.2	\\
23.123	&	OH	&	X1.5	&	11	&	$A'$	&		&		&	22.9	&	$\pm$	&	0.8	&	6.1	&	$\pm$	&	1.2	\\
23.205	&	OH	&	X0.5	&	11	&	$A''$	&		&		&	9.9	&	$\pm$	&	0.9	&	6.0	&	$\pm$	&	1.2	\\
23.243	&	OH	&	X0.5	&	11	&	$A'$	&		&		&	19.4	&	$\pm$	&	0.8	&	6.0	&	$\pm$	&	1.2	\\
24.512	&	[Fe II]	&	${}^4 F_{5/2}-{}^4 F_{7/2}$	&		&		&		&		&	71.6	&	$\pm$	&	0.3	&	5.4	&	$\pm$	&	1.0	\\
25.035	&	OH	&	X1.5	&	10	&	$A''$	&		&		&	12.3	&	$\pm$	&	0.3	&	5.2	&	$\pm$	&	1.0	\\
25.090	&	OH	&	X1.5	&	10	&	$A'$	&		&		&	32.1	&	$\pm$	&	0.3	&	5.2	&	$\pm$	&	1.0	\\
25.216	&	OH	&	X0.5	&	10	&	$A''$	&		&		&	15.5	&	$\pm$	&	0.3	&	5.2	&	$\pm$	&	1.0	\\
25.251	&	[S I]	&	${}^3 P_1 - {}^3 P_2$	&		&		&		&		&	77.9	&	$\pm$	&	0.3	&	5.2	&	$\pm$	&	1.0	\\
25.258	&	OH	&	X0.5	&	10	&	$A'$	&		&		&	39.9	&	$\pm$	&	8.0	&	5.2	&	$\pm$	&	1.0	\\
25.989	&	[Fe II]	&	${}^6 F_{7/2}-{}^6 F_{9/2}$	&		&		&		&		&	268.8	&	$\pm$	&	0.5	&	4.9	&	$\pm$	&	0.9	\\
    \enddata
\tablenotetext{3}{For OH, electronic configuration; X1.5 = $X^2\Pi_{3/2}$ and X0.5 = $X^2\Pi_{1/2}$. For $\rm CO_2$, vibrational quantum numbers $\nu_1 \nu_2^l \nu_2$ of initial and final state. For $\rm H_2$, vibrational quantum number. For ions and atoms, initial and final electron configuration in $^{2S+1}L_J$ form.}
\tablenotetext{4}{For OH, lower-state orbital angular momentum quantum number $N$.  For $\rm CO_2$, rotational transition identification, with P-Q-R signifying $\Delta J = +1,0,-1$, and a numeral for lower-state total angular momentum quantum number $J$.}
\tablenotetext{5}{For OH, symmetry $(A')$ or antisymmetry $(A'')$ of wavefunction about molecular orbital plane.}
\tablenotetext{6}{"+" indicates a transition spectrally unresolved from the following one. Its fluxes (7) and extinction factors (8) apply to the unresolved combination. }
\tablenotetext{7}{Line flux $\pm$ 1$\sigma$ uncertainty, or $3\sigma$ upper limit.}
\tablenotetext{8}{Extinction correction factors and their uncertainties, derived as per Section \ref{extinction}. The factor is multiplied with the observed flux, and the uncertainty propagated with that of the observed flux, to produce extinction-corrected values in the analysis.}
\tablecomments{Lines not listed are not detected.  Upper limits may be calculated by interpolation among the entries of columns 1 and 7.}
\end{deluxetable*}

\clearpage
\bibliography{biblio}{}  

\end{document}